\documentclass[aps,twocolumn,superscriptaddress]{revtex4-1}

\usepackage[utf8]{inputenc}
\usepackage[english]{babel}
\usepackage{amsmath}
\usepackage{amsfonts}
\usepackage{amssymb}
\usepackage{makeidx}
\usepackage{graphicx}
\usepackage{lmodern}
\usepackage{kpfonts}
\usepackage[left=1.5cm,right=1.5cm,top=2cm,bottom=2cm]{geometry}

\usepackage{xcolor}
\usepackage{hyperref}
\hypersetup{linkbordercolor=cyan}

\usepackage{amsmath}
\newcommand*\diff{\mathop{}\!\mathrm{d}}

\usepackage{verbatim}
\usepackage{caption}
\usepackage{subcaption}
\usepackage{tabularx}
\usepackage{floatrow}

\usepackage{sidecap}

\usepackage[margin=0.2cm]{subcaption}
\usepackage[toc,page]{appendix}
\usepackage{xcolor} 
\usepackage{graphicx}
\usepackage[export]{adjustbox}
\graphicspath{
{C:/Users/Utilisateur/Documents/Uni/these/latex/images/}
{C:/Users/Utilisateur/Dropbox/images/}
{C:/Users/Utilisateur/Dropbox/images/Langer/isotropic_annihilation/}
{C:/Users/Utilisateur/Dropbox/images/Langer/boundary_annihilation/}
{C:/Users/Utilisateur/Dropbox/images/Langer/2skyrmions_annihilation/}
{C:/Users/louise/Dropbox/images/}
{C:/Users/louise/Dropbox/images/screencaps/}
{C:/Users/louise/Dropbox/images/Langer/isotropic_annihilation/}
{C:/Users/louise/Dropbox/images/Langer/boundary_annihilation/}
{C:/Users/louise/Dropbox/images/Langer/2skyrmions_annihilation/}
{C:/Users/louise/Dropbox/images/Langer/}
{C:/Users/louise/Dropbox/images/Langer/non_mag_defect/}
{C:/Users/louise/Dropbox/images/Langer/blochsk_collapse/}
{C:/Users/louise/Dropbox/images/Langer/antisk_collapse/}
{/home/desplat/Dropbox/images/Langer/boundary_annihilation/}
{/home/desplat/Dropbox/images/Langer/isotropic_annihilation/}
{/home/desplat/Dropbox/images/Langer/2skyrmions_annihilation/}
{/home/desplat/Dropbox/writing/latex/images/}
{/home/desplat/Dropbox/images/Langer/}
{/home/louise/Dropbox/writing/latex/images/}
{/home/louise/Dropbox/images/}
{/home/louise/Dropbox/images/screencaps/}
{/home/louise/Dropbox/images/Langer/isotropic_annihilation/}
{/home/louise/Dropbox/images/Langer/boundary_annihilation/}
{/home/louise/Dropbox/images/Langer/2skyrmions_annihilation/}
{/home/louise/Dropbox/images/Langer/antisk_collapse/}
{/home/louise/Dropbox/images/Langer/blochsk_collapse/}
{/home/louise/Dropbox/images/Langer/non_mag_defect/}
{/home/louise/Dropbox/images/Langer/}
{/home/louise/Dropbox/images/Langer/curved_boundary/}
}

\usepackage{multirow}

\begin{document}

\title{Thermal stability of metastable magnetic skyrmions: \\Entropic narrowing and significance of internal eigenmodes}%

\author{L. Desplat}%
\email{l.desplat.1@research.gla.ac.uk}
\affiliation{SUPA School of Physics and Astronomy, University of Glasgow, G12 8QQ Glasgow, United Kingdom}
\affiliation{Centre for Nanoscience and Nanotechnology, CNRS, Université Paris-Sud, Université Paris-Saclay, 91120 Palaiseau, France}

\author{ D. Suess}
\affiliation{Christian Doppler Laboratory, Physics of Functional Materials, Faculty of Physics, University of Vienna, 1090 Vienna, Austria}

\author{J-V. Kim}
\affiliation{Centre for Nanoscience and Nanotechnology, CNRS, Université Paris-Sud, Université Paris-Saclay, 91120 Palaiseau, France}

\author{ R. L. Stamps}
\affiliation{Department of Physics and Astronomy, University of Manitoba, Winnipeg, Manitoba, R3T 2N2 Canada}
\affiliation{SUPA School of Physics and Astronomy, University of Glasgow, G12 8QQ Glasgow, United Kingdom}

\date{\today}

\begin{abstract}
We compute annihilation rates of metastable magnetic skyrmions using a form of Langer's theory in the intermediate-to-high damping (IHD) regime. For a Néel skyrmion, a Bloch skyrmion, and an antiskyrmion, we look at two possible paths to annihilation: collapse and escape through a boundary. We also study the effects of a curved vs. a flat boundary, a second skyrmion and a non-magnetic defect.
We find that the skyrmion's internal modes play a dominant role in the thermally activated transitions compared to the spin-wave excitations and that the relative contribution of internal modes depends on the nature of the transition process. 
Our calculations for a small skyrmion stabilized at zero-field show that 
collapse on a defect is the most probable path. In the absence of a defect,
the annihilation is largely dominated by escape mechanisms, even though in this case the activation energy is higher than that of collapse processes.  Escape through a flat boundary is found more probable than through a curved boundary. The potential source of stability of metastable skyrmions is therefore found not to lie in high activation energies, nor in the dynamics at the transition state, but comes from entropic narrowing in the saddle point region which leads to lowered attempt frequencies. This narrowing effect is found to be primarily associated with the skyrmion's internal modes. 
\end{abstract}

\maketitle


\section{Introduction}

\paragraph*{}Magnetic skyrmions are localized, topologically non-trivial solitonic magnetic textures stabilized by competing isotropic and anisotropic exchange couplings, such as the Dzyaloshinskii-Moriya interaction (DMI) \cite{dzyaloshinskii,moriya}. Chiral skyrmion solutions were theoretically investigated in the 1990s for thermodynamically stable \cite{bogdanov1989thermodynamically,bogdanov1994thermodynamically} and metastable \cite{ivanov1990magnetic} configurations. Isolated skyrmions exist as metastable excitations of the ferromagnetic ground state and can be long-lived.  The computation of accurate lifetimes for isolated skyrmions is challenging since the decay rate of metastable states depends on details of the fluctuations about stable and unstable configurations as well as the activation barrier.  In recent years, skyrmions have attracted interest for potential spintronic applications as racetrack memories and logic gates \cite{fert2013skyrmions}. In order to be used in viable room-temperature technology devices, individual skyrmion bits need to be highly stable in a  wide range of temperatures. For this reason, precisely estimating  and understanding their stability against thermal fluctuations is a crucial step to designing metastable states with long lifetimes.

\begin{figure}[hbtp]
		\includegraphics[width=.5\textwidth]{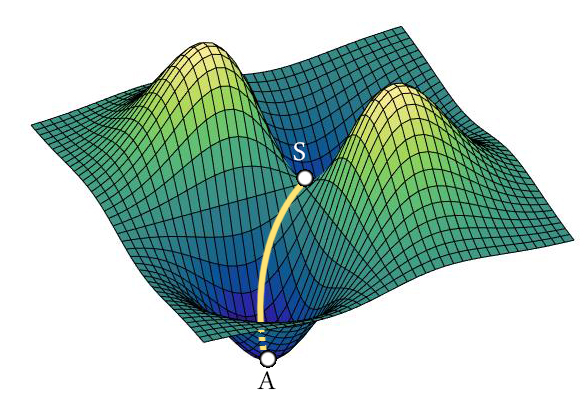}
		\caption{Typical energy surface of a system possessing a metastable local minimum $A$ and a stable global minimum $M$ separated by a saddle point $S$. The reaction coordinate is represented by a yellow line and corresponds to the path of minimun energy connecting $A$, $S$ and $M$.}
	\label{fig:3d_landscape}
\end{figure}
	
\paragraph*{} An individual skyrmion state $A$ is separated from the ferromagnetic ground state $M$ by an activation energy (see Fig. \ref{fig:3d_landscape}).
The activation energy $\Delta E = E_S - E_A$ corresponds to the height of the barrier that needs to be overcome by the individual skyrmion to reach the saddle point (SP) $S$ during a transition process. If several saddle points are present in the energy landscape, the total escape rate out of the metastable well is the sum of the escape rates over each saddle point. For a given mechanism, the path through the energy landscape that connects $A$, $S$ and $M$ is referred to as the reaction coordinate. In the case of multidimensional systems, the most favorable path typically involves a first order saddle point, which corresponds to a local minimum in the energy with respect to all degrees of freedom except one: the reaction coordinate, which is associated with a local maximum. At finite temperature, the magnetization is coupled to the environment which acts as a heat reservoir of constant temperature $T$ and leads to fluctuations of the magnetic moments. Over time, rare energy fluctuations in excess of the barrier height may promote the skyrmion state to the transition state. From there, the system may spend some time at the barrier top in a superposition of a large number of modes of stable fluctuations. There exists, however, an unstable mode that eventually provides a means to overcome the barrier and reach the ferromagnetic ground state. The decay rate measures the average frequency for that series of events and therefore gives an estimate of a skyrmion's stability.

\paragraph*{}In the present work, we apply Langer's theory for the decay of a metastable state \cite{langer} to the problem of individual skyrmion annihilation. The theory constitutes the most complete treatment of the extension of Kramers theory to a multidimensional phase space in the intermediate-to-high damping (IHD) regime \cite{coffey}.  The extension to many degrees of freedom allows the theory to be applied to magnetic spin systems with energies determined by exchange and dipole-dipole coupling, and can therefore be used to assess the stability of individual skyrmions. The restriction to the IHD regime means the scope of the theory for magnetic systems is limited to cases where the precessional dynamics can be neglected, in the sense that it does not impact significantly the transition path, and the time-scale of the transition is set by the dissipation rate. The energy barrier must be high compared to thermal energy, typically $\Delta E \sim 5k_B T $ \cite{coffey} so that the system remains close to equilibrium at all times. This also ensures that barrier re-crossing events are negligible. We therefore  consider the rate of skyrmion nucleation from the ferromagnetic ground state to be zero. The rate of decay is given by an Arrhenius-type law,
\begin{equation}\label{eq:escaperate}
\Gamma = \Gamma_0 e^{-\Delta E /k_B T}.
\end{equation}
 The prefactor $\Gamma_0 $ corresponds to a fundamental fluctuation rate and is linked to characteristic time scales of the dynamics of the barrier-crossing. Given the above hypotheses, it is defined as,
 \begin{equation}\label{eq:prefactor}
\Gamma_0 =\frac{\lambda +}{2 \pi} \Omega_0,
\end{equation}
in which $\Omega_0$ is the ratio of energy curvatures in the metastable well and at the saddle point, and $\lambda_+$ is a prefactor that takes into account the dissipative dynamics of the system at the top of the barrier 
\cite{coffey}.
The meaning and derivation of these terms for magnetic spin systems are discussed later in the text. It is important to note the presence of the exponential in Eq. (\ref{eq:escaperate}), which shows that the decay of metastable states takes place over  time-scales which are much longer than the time scales linked with the intrinsic dynamics of the system. For this reason, solely understanding the dynamics is not enough in order to predict the processes by which skyrmions  annihilate, and it is essential to study and understand the annihilation mechanisms themselves. For systems with many degrees of freedom and many-body interactions such as magnetic spin systems, this is often an arduous task and relying on numerical schemes becomes almost unavoidable. The difficulty in computing transition rates for such a class of systems thus lies in the identification of the first order saddle point(s) in the energy landscape on the one hand, and on the correct evaluation of the different terms in the rate prefactor on the other hand.

 \paragraph*{}To date, activation energies of individual skyrmions in two-dimensional systems were calculated via the geodesics nudged elastic bands (GNEB) \cite{gneb} scheme \cite{uzdin2017effect,lobanov2016mechanism,von2017enhanced}, Monte Carlo simulations \cite{hagemeister2015stability} and experimental investigations \cite{wild2017entropy}, including estimations of the Arrhenius prefactor \cite{wild2017entropy,hagemeister2015stability}. Two main annihilation mechanisms, namely istropic collapse and escape through a boundary, were previously reported in \onlinecite{uzdin2017effect,lobanov2016mechanism}. Ref.  \onlinecite{wild2017entropy,hagemeister2015stability} also discussed the importance of entropic effects on skyrmions' stability. More recently, Bessarab \textit{et al.} \cite{bessarab2018annihilation} computed average lifetimes of racetrack skyrmions stabilized at high magnetic fields using harmonic transition state theory \cite{bessarab2012htst} and with respect to the two mechanisms mentioned above. An semi-infinite racetrack was simulated by assuming periodic boundary conditions along one direction, which gives rise to translational invariance with respect to the skyrmion position. The subsequent treatment of Goldstone modes then yields a temperature dependence of the rate prefactor, as well as a sample width dependence for escape mechanisms. In the present work, we stick to a finite-sized system and skyrmions stabilized at zero-field. The eigenfrequencies associated to the translational modes are found not to be numerical zeros, and are therefore not treated as Goldstone modes.
 
A previous implementation of Langer's theory was done by Fiedler \textit{et al.} \cite{seuss} based on the finite element method and applied to obtain the attempt frequencies in a small ferromagnetic cube and a graded media grain. However, in the micromagnetic framework and within a three-dimensional world, magnetic skyrmions typically decay via the formation of a Bloch point, a topological singularity where the continuity of magnetism is broken \cite{rohart2016path}. While Bloch points do not strictly exist in two dimensions, equivalent processes in 2D have been reported (see discussion in \cite{rohart2016path} and subsequent Ref.). The use of atomistic simulations therefore seems necessary in order to avoid a mesh-size dependency of the activation rates \cite{suess2017repulsive}.

\paragraph*{}The paper is organized as follows. In Sec. \ref{sec:mecha}, we firstly present different decay mechanisms of an individual skyrmion  stabilized at zero-field: collapse of an isolated skyrmion \cite{uzdin2017effect,lobanov2016mechanism,von2017enhanced,bessarab2018annihilation} and the effect of a defect \cite{uzdin2017effect} and of a second skyrmion on the collapse, and escape through a flat boundary \cite{uzdin2017effect,lobanov2016mechanism,bessarab2018annihilation} as well as through a curved boundary. The escape and collapse mechanisms are studied for a Néel skyrmion, a Bloch skyrmion, and an antiskyrmion. In Sec. \ref{sec:prefactor}, we give some details on the calculation of the terms of the rate prefactor. Finally, in Sec. \ref{sec:results}, annihilation rates are calculated and we discuss the role of the internal eigenmodes of skyrmions in the annihilation as well as the meaning behind the obtained attempt frequencies and the source of potential stability of individual skyrmions.

\section{Annihilation mechanisms and activation barriers}\label{sec:mecha} Our system is a simple bidimensional square lattice of $N$ magnetic spins $\{\hat{m}_i\}$ with a constant magnitude that we set to unity, and we assume open boundary conditions.
The corresponding Heisenberg-type Hamiltonian is
\begin{equation}\label{eq:hamiltonian}
	E = - J_{ex}\sum_{<ij>} \hat{m}_i \cdot \hat{m}_j - \sum_{<ij>} \vec{D}_{ij} \cdot \big( \hat{m}_i \times \hat{m}_j \big)  -  K \sum_i m_{z,i}^2,
	\end{equation}	
where  $J_{\text{ex}}$ is the strength of the isotropic Heisenberg exchange, $D_{ij}$ is the Dzyaloshinskii vector between sites $i$ and $j$, and $K$ is the perpendicular, uniaxial anisotropy constant. Different types of topological defects are stabilized by changing the underlying symmetry of the DMI. The summations over $<ij>$ are performed over first nearest-neighbor pairs.  Minimum energy paths (MEPs) in the energy landscape are computed using our implementation of the GNEB method \cite{gneb} with a climbing image (CI) scheme \cite{climbingimage} to identify saddle points with high accuracy. Successive states of the system along the path are referred to as images (im.). The chosen parameters and further details on the simulations are given in Appendix \ref{app:simulations}. We stress that these mechanisms correspond to most probable paths in the energy surface and do not take dynamics into account. For each of them, we make sure we indeed find a MEP by checking that there is only a single negative curvature at the saddle point.


\begin{figure}[hbtp]
	\begin{subfigure}[t]{1\textwidth}
		\includegraphics[width=1\textwidth]{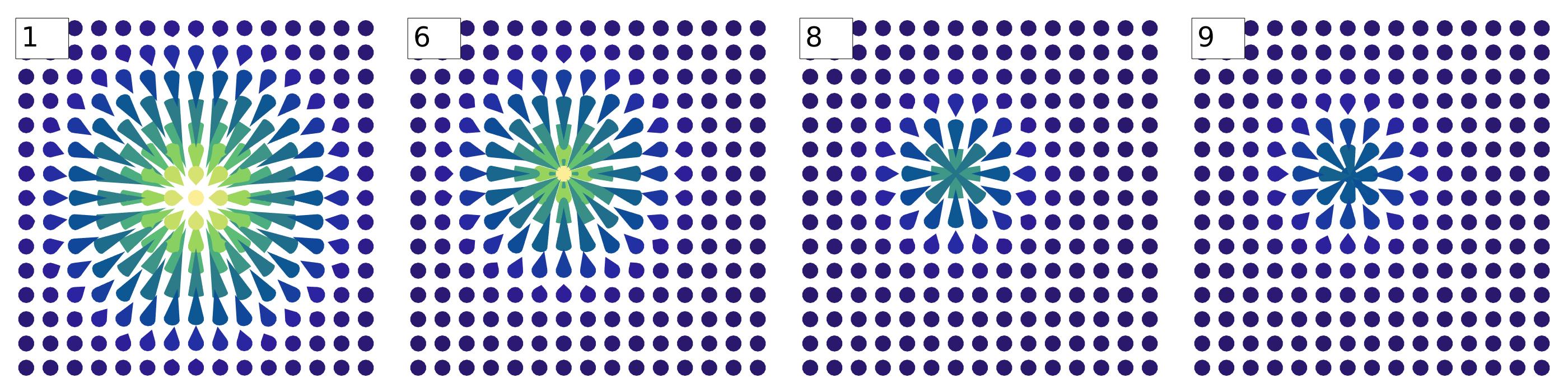}
		\caption{}
		\label{fig:iso_mechanism}
	\end{subfigure}\hfill
		
	\begin{subfigure}[t]{1\textwidth}
		\includegraphics[width=1\textwidth]{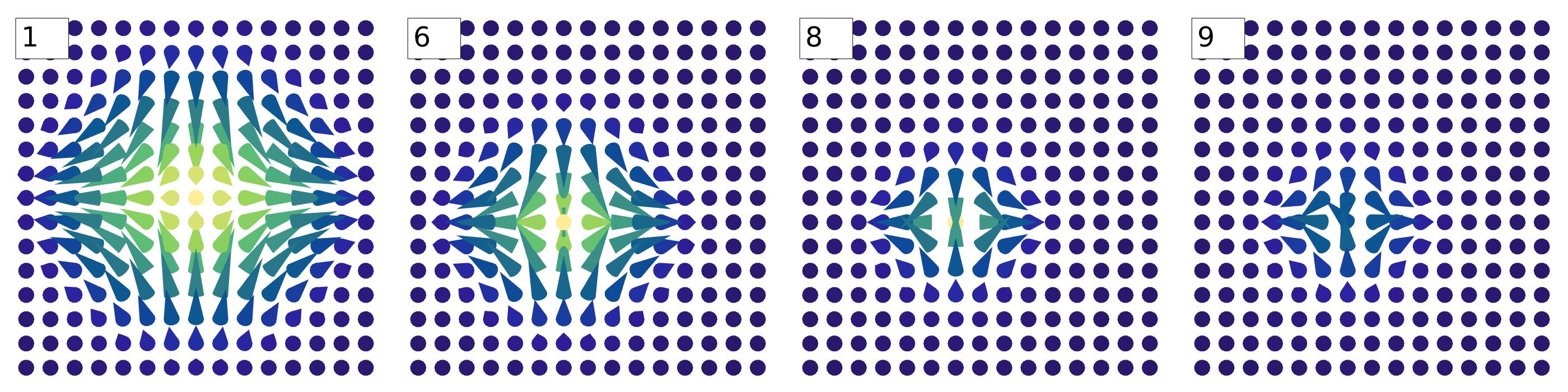}
		\caption{}
		\label{fig:antisk_mechanism}
	\end{subfigure}\hfill
	
		\begin{subfigure}[t]{1\textwidth}
		\includegraphics[width=1\textwidth]{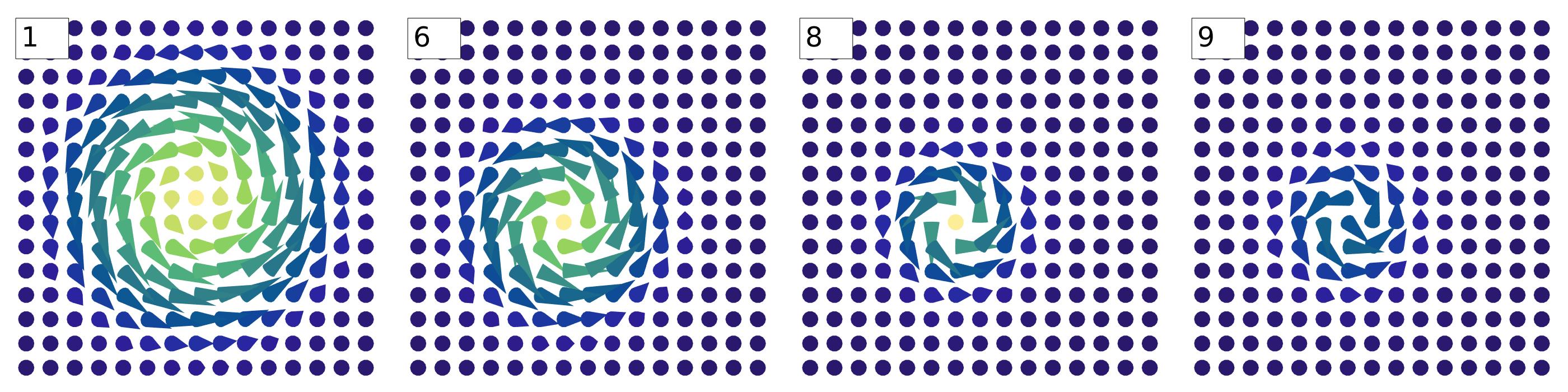}
		\caption{}
		\label{fig:blochsk_mechanism}
	\end{subfigure}\hfill
	
			\begin{subfigure}[t]{1\textwidth}
		\includegraphics[width=1\textwidth]{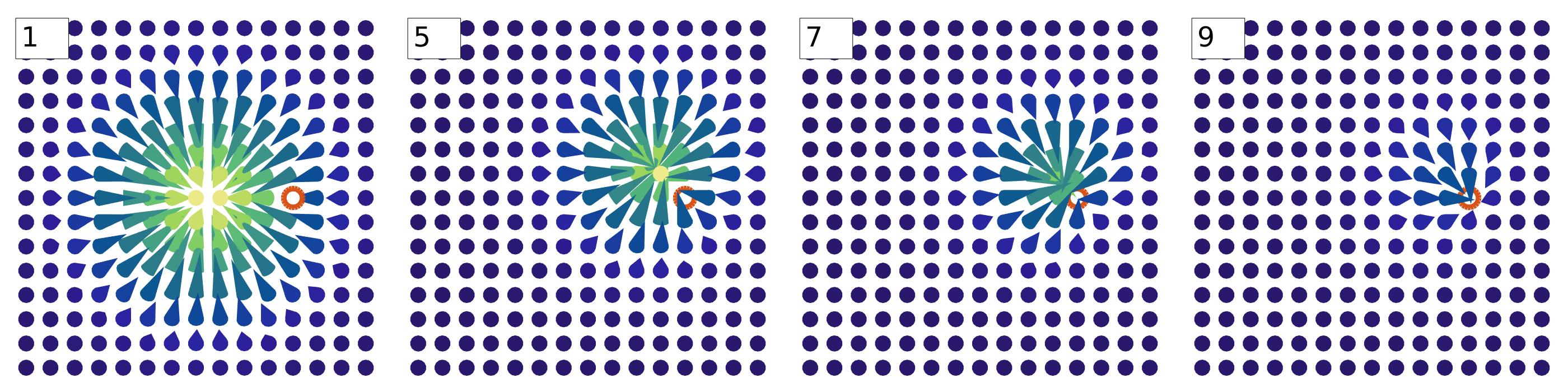}
		\caption{}
		\label{fig:defect_mechanism}
	\end{subfigure}\hfill
	
	\begin{subfigure}{1\textwidth}
		\includegraphics[width=1\textwidth]{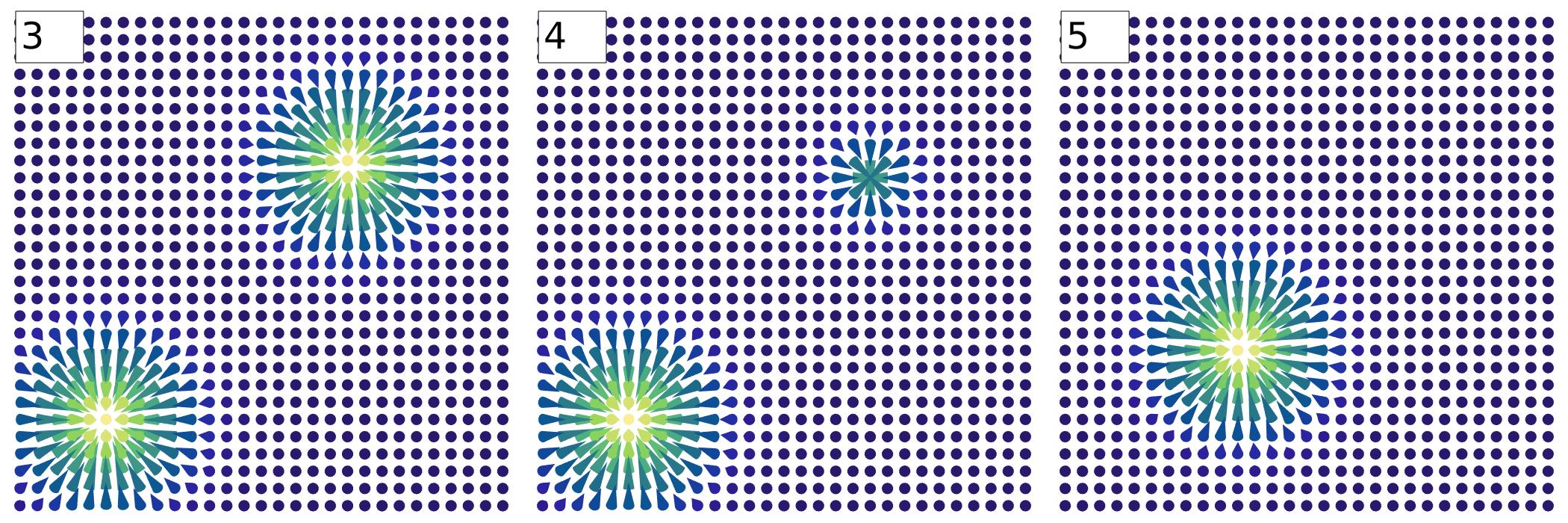}
		\caption{}
	\label{fig:2sk_mechanism}
	\end{subfigure}\hfill	
	\begin{subfigure}[t]{.2\textwidth}
				\includegraphics[width=1\textwidth]{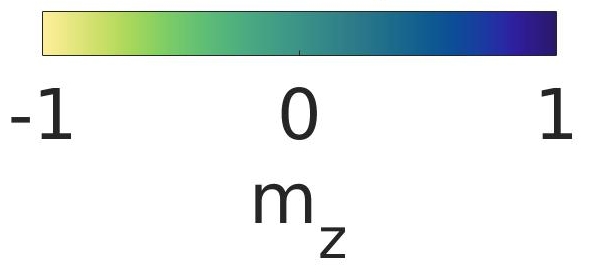}
	\end{subfigure}

	\caption{Spin maps of the collapse mechanisms. The index in the top left-hand corner corresponds to the image index of the GNEB method. The saddle point corresponds to the state preceding the flipping of the core spin. 
(a) Néel skyrmion. (b) Antiskyrmion. (c) Bloch skyrmion. (d) Néel skyrmion in the presence of a non-magnetic defect. (e) Néel skyrmion in the presence of a second skyrmion.
\label{fig:mechanisms_all}}		
\end{figure}


\begin{figure}[hbtp]
		\begin{subfigure}[t]{1\textwidth}
		\includegraphics[width=1\textwidth]{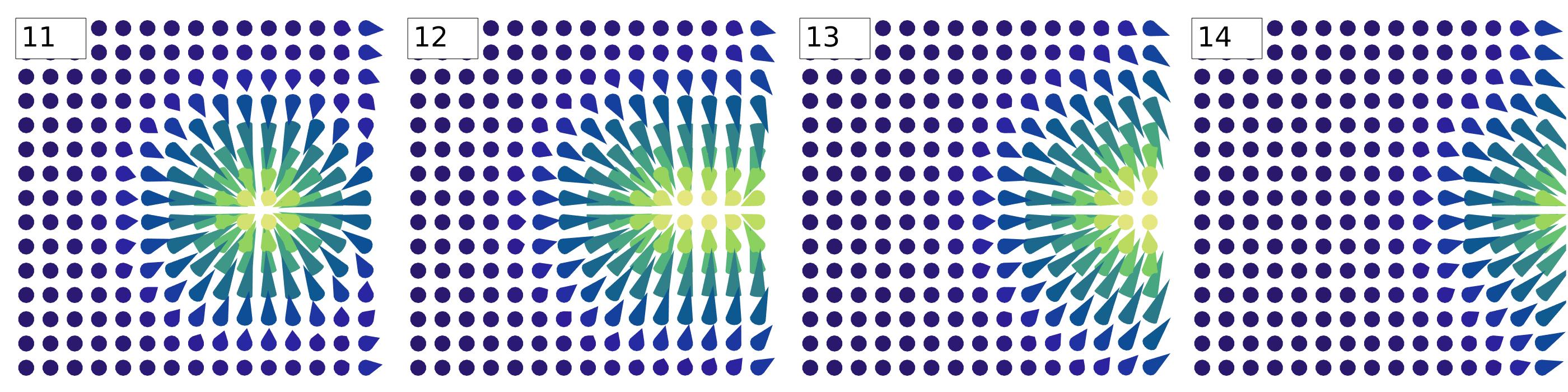}
		\caption{}
		\label{fig:boundary_neel}
	\end{subfigure}
		\begin{subfigure}[t]{1\textwidth}
		\includegraphics[width=1\textwidth]{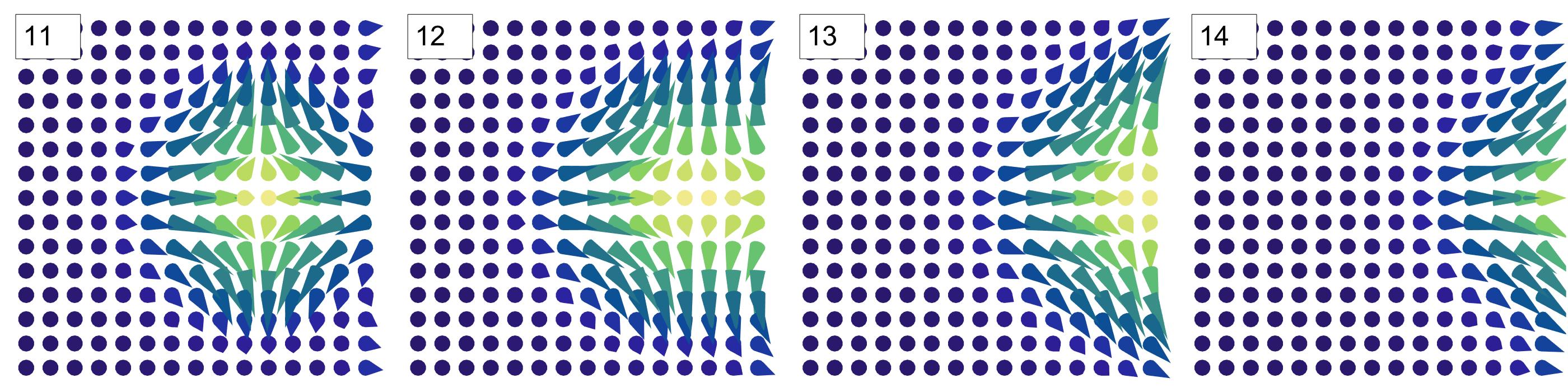}
		\caption{}
		\label{fig:boundary_antisk}
	\end{subfigure}	
			\begin{subfigure}[t]{1\textwidth}
		\includegraphics[width=1\textwidth]{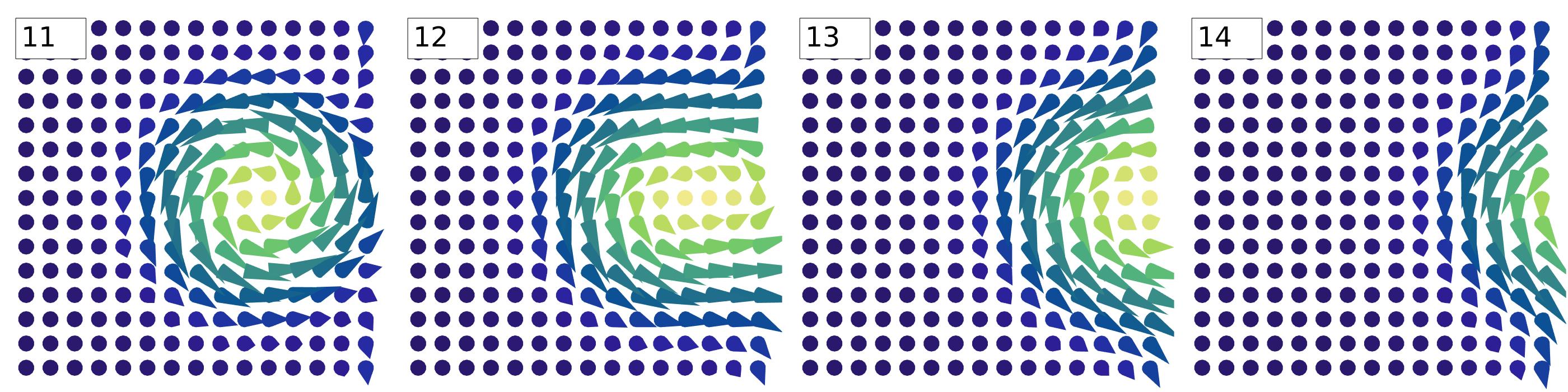}
		\caption{}
		\label{fig:boundary_bsk}
	\end{subfigure}
			\begin{subfigure}[t]{1\textwidth}
		\includegraphics[width=1\textwidth]{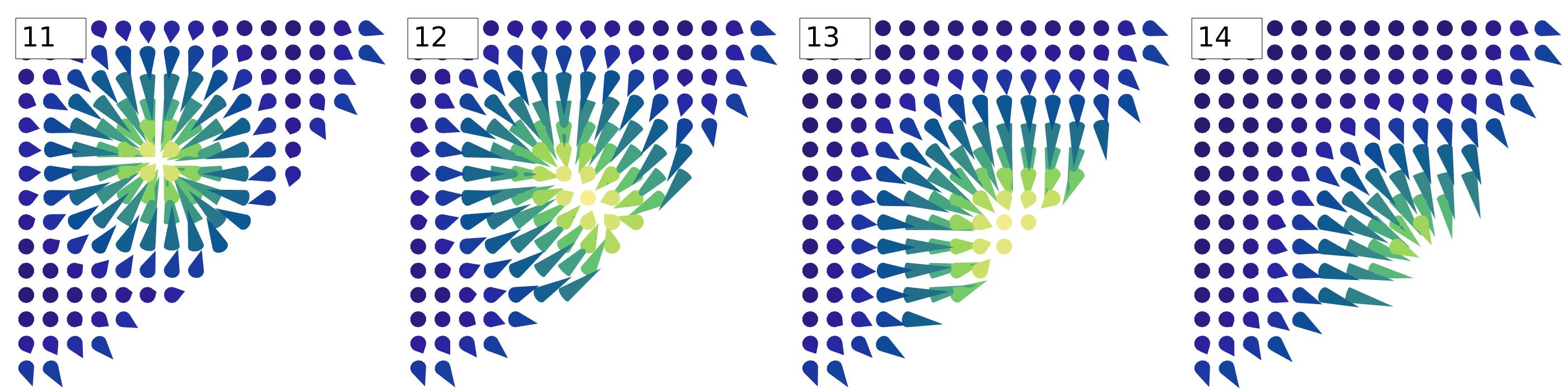}
		\caption{}
		\label{fig:curved_boundary_mechanism}
	\end{subfigure}

	\begin{subfigure}[t]{.2\textwidth}
				\includegraphics[width=1\textwidth]{cm.jpg}
	\end{subfigure}
		\caption{Spin maps of the escape mechanisms. The conventions are the same as in Fig. \ref{fig:mechanisms_all}. On all the subfigures, the saddle point is im. 11, which is the state where the skyrmion sits tangent to the boundary. (a) Néel skyrmion. (b) Antiskyrmion. (c) Bloch skyrmion. (d) Néel skyrmion escaping through a curved boundary. \label{fig:boundary_mecha_all}}		
\end{figure}


\begin{figure*}[hbtp]
\centering	
	\begin{subfigure}[t]{.4\textwidth}
		\adjincludegraphics[width=1\textwidth,trim={0 {.255\height} {.26\width} {.04\height}},clip]{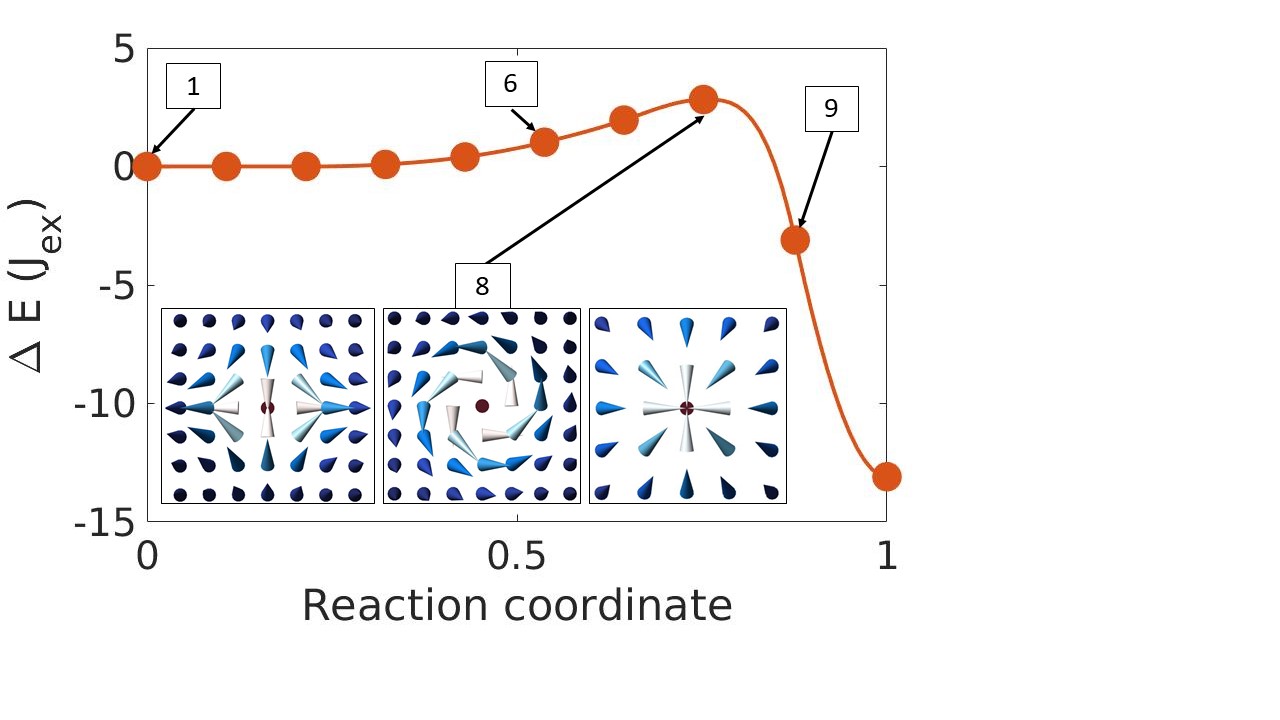}
\caption{}
	\label{fig:Etot_iso}
	\end{subfigure}	
	\begin{subfigure}[t]{.4\textwidth}
		\adjincludegraphics[width=1\textwidth,trim={0 {.255\height} {.26\width} {.04\height}},clip]{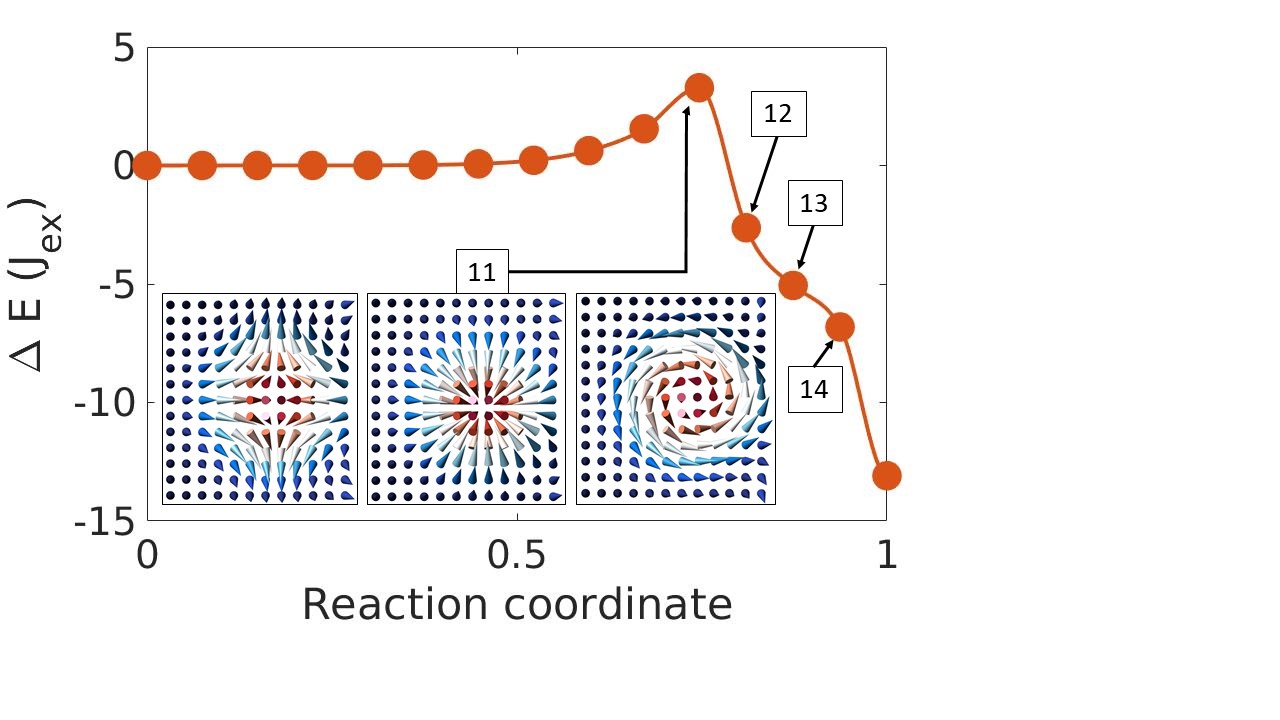}
		\caption{}
					\label{fig:Etot_boundary}
	\end{subfigure}\hfill	
	
			\begin{subfigure}[t]{.1\textwidth}
				\includegraphics[width=1\textwidth]{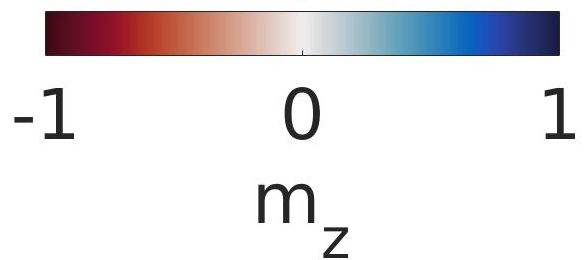}
	\end{subfigure}
	
			\begin{subfigure}[t]{.4\textwidth}
		\adjincludegraphics[width=1\textwidth,trim={0 {.1\height} {.26\width} {.04\height}},clip]{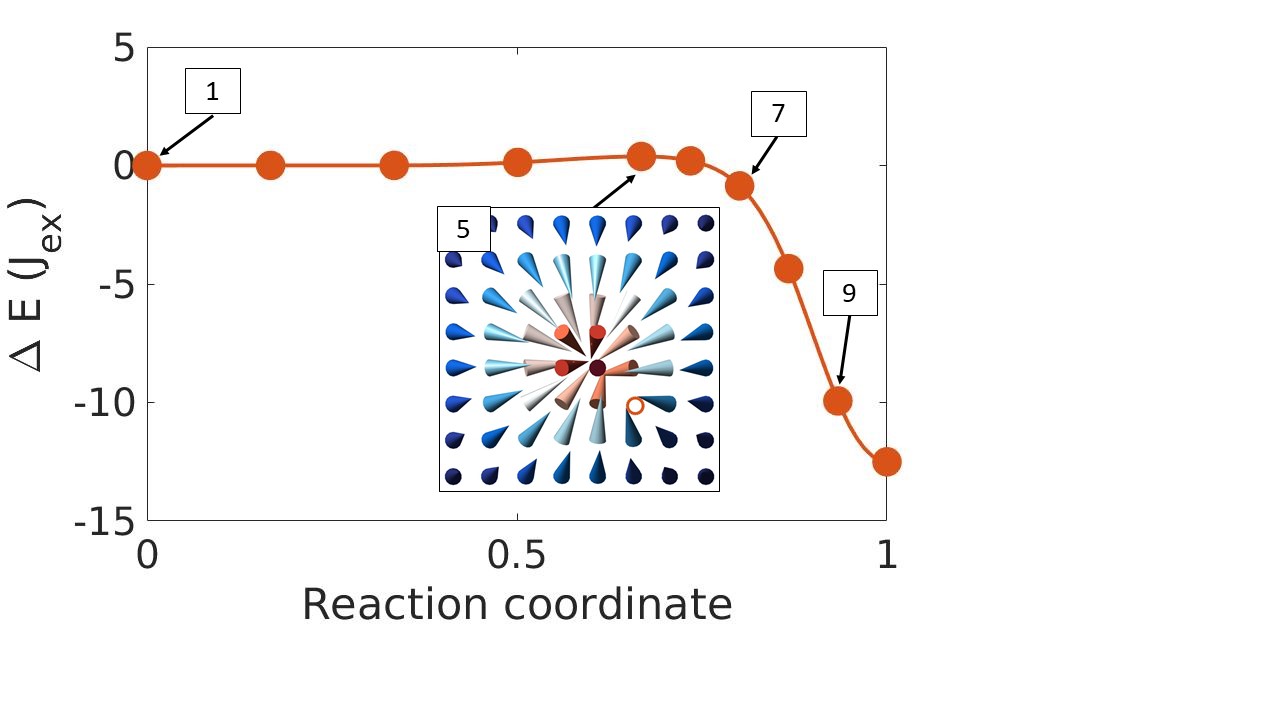}
		\caption{}
		\label{fig:Etot_def}
	\end{subfigure}
	\begin{subfigure}[t]{.4\textwidth}
		\adjincludegraphics[width=1\textwidth,trim={0 {.1\height} {.26\width} {.04\height}},clip]{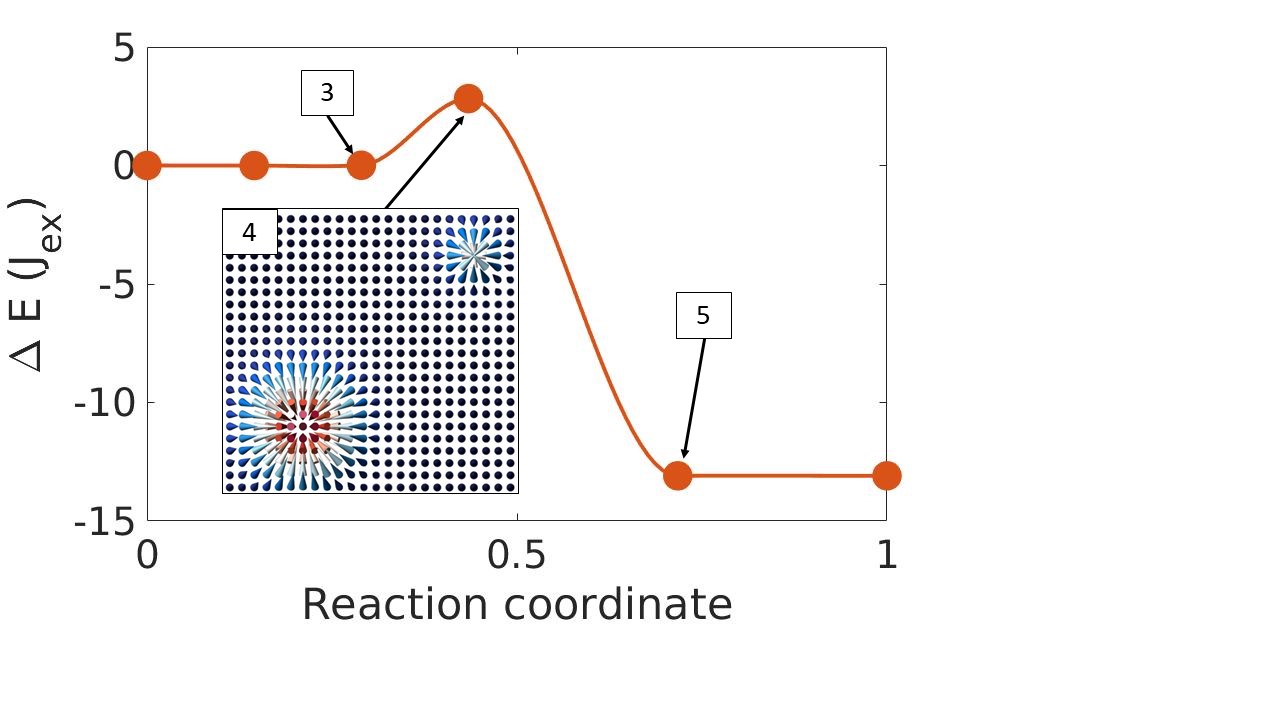}
		\caption{}
		\label{fig:Etot_2sk}
	\end{subfigure}

\caption{Interpolated energy profiles along the normalized reaction coordinate. Each dot corresponds to an image of the system on the energy surface. The insets show a closeup of the spin configuration at the saddle point. 
  (a) Collapse of a single skyrmion. The energy increases slowly as the skyrmion shrinks (im. [1-8]). Past the barrier top (im. 8), it annihilates by breaking the radial symmetry which is accompanied by a brutal decrease in the energy. (b) Escape through a flat boundary. The energy rises as the skyrmion gets closer to the edge (im. [9 - 11]). The top of the barrier is the state tangent to the boundary (im. 11). Past that point, the skyrmion disappears through the edge. This is accompanied by a rapid drop in the energy, with a notable slowdown halfway through the process as half the skyrmion has disappeared (im. 13). (c) Collapse  in the presence of a single non-magnetic defect. The skyrmion shrinks in size as the core moves towards the defect. This costs very little energy (im [1-5]). Past the saddle point, the skyrmion collapses on the defect. (d) Collapse in the presence of another skyrmion. The skyrmions get closer to each other, which at first costs little energy (im. [1-3]) until a critical distance is reached where the collapse of the upper skyrmion is initiated. The saddle point is the same as that of the first mechanism for the upper skyrmion, while the other remains stable (im. 4).}
\label{fig:Etot_all}
\end{figure*}

 \subsection{Collapse of a skyrmion.} 
  \paragraph{Isolated skyrmion.}
We first study the case of the collapse of a single individual skyrmion on itself. Key steps in the process are shown in Fig. \ref{fig:iso_mechanism} for the Néel skyrmion, and the corresponding energy profile is found in Fig. \ref{fig:Etot_iso}. The skyrmion progressively shrinks onto itself without breaking cylindrical symmetry. This is accompanied by a slow increase in energy. The critical fluctuation, which corresponds to the state preceding the flipping of the core spin, constitutes the saddle point configuration (im. 8 on Fig. \ref{fig:iso_mechanism}). Once the core begins to reverse, symmetry is broken: the remaining spins flip, the energy drops dramatically and the system overcomes the saddle and reaches the ferromagnetic ground state. The energy profile shown on Fig. \ref{fig:Etot_iso} appears similar to the ones previously reported in Ref.  \onlinecite{lobanov2016mechanism}, in which the GNEB scheme was also used.  The collapse of an antiskyrmion and a Bloch skyrmion are shown on Fig. \ref{fig:antisk_mechanism} and \ref{fig:blochsk_mechanism}, respectively. They both exhibit a very similar behaviour to that of the Néel skyrmion, with a breaking of the symmetry past the saddle point corresponding to the flipping of the core. The energy profile along the path is the same for all three types of skyrmions, and the activation energy is found to be $\Delta E_{\text{col}} =$ 2.83 $J_{\text{ex}}$ ($\sim 10 k_B T$ at 300K with our choice of $J_{\text{ex}}$).

 \paragraph{Effect of a non-magnetic defect.}
The effect of a single non-magnetic defect on the collapse was studied. Im. 1 on Fig. \ref{fig:defect_mechanism} shows that the skyrmion relaxes such that the defect sits where the spins lie in-plane. As the core moves towards the defect (here it does so diagonally), the skyrmion shrinks in size. This all costs very little energy, as seen on the energy profile of Fig \ref{fig:Etot_def}. The saddle point is im. 5 on the path and corresponds to a bigger skyrmion compared to the case with no defect, but the skyrmion is deformed and rendered asymmetric by the defect. Consistently with results presented in Ref. \onlinecite{uzdin2017effect}, the presence of the defect significantly lowers both annihilation and nucleation barriers, and we obtain  $\Delta E_{\text{def}}$ = 0.38 $  J_{\text{ex}}$, which is one order of magnitude lower than $\Delta E_{\text{col}}$. 

 \paragraph{Effect of a second skyrmion.}
We consider what happens when two skyrmions approach one another. When the skyrmion cores are initally aligned along $X$ or $Y$, we observe that they rotate in order to approach each other along the lattice diagonal. Presumably, this is more energically favorable due to the choice of first neighbor interactions on the square lattice. Consequently, we simply initialize the skyrmions diagonally from each other. We set the transition path for a merging of the two skyrmions into one, as observed experimentally in the case of the decay of a skyrmion lattice into the helical state \cite{wild2017entropy}. However, the search for a first order SP consistently results in a switch in mechanism, and the collapse of one of the skyrmions is relaxed instead of the merging (Fig. \ref{fig:2sk_mechanism}). With the present set of parameters, this might hint at the fact that the merging mechanism involves a higher order SP, and is therefore less favorable than the collapse. The skyrmions get closer to each other, which at first costs almost no energy (im. [1-3] in Fig. \ref{fig:Etot_2sk}) until a critical distance is reached where the collapse of the upper skyrmion is initiated. The saddle point is the same as that of the first mechanism for the upper skyrmion, while the other one remains stable (im. 4 on Fig. \ref{fig:2sk_mechanism}). 
The activation energy for collapse remains the same: $\Delta E_{\text{2sk}} = 2.82 J_{\text{ex}}$. 


\subsection{Escape through a boundary.} 
\paragraph{Flat boundary.}
Escape through a boundary is another possible path in finite-sized systems. The canting of the spins along the edge in the presence of DMI makes the boundary repulsive, such that the total energy increases as the skyrmion leaves the center of the lattice and moves towards an edge (Fig. \ref{fig:Etot_boundary}). For the Néel skyrmion, the antiskyrmion and the Bloch skyrmion, the saddle point corresponds to a position where the skyrmion sits tangent to the boundary (im. 11 on Fig. \ref{fig:boundary_neel}, \ref{fig:boundary_antisk} and \ref{fig:boundary_bsk}, respectively), as is also shown in Ref. \onlinecite{uzdin2017effect}. Past the saddle point, the skyrmion deforms and elongates as it comes in contact with the edge and begins to disappear. This is accompanied by a large decrease in the energy. Im. 13 on Fig. \ref{fig:boundary_neel}, \ref{fig:boundary_antisk}, \ref{fig:boundary_bsk}, and \ref{fig:Etot_boundary} corresponds to a half-skyrmion sitting on the edge. In the vicinity of this point, the decrease in energy appears to slow down, before speeding up again as the rest of the remaining skyrmion disappears. The activation energy obtained for this mechanism is the highest one of the three processes studied here, although they are all of a similar magnitude of 2-3$J_{\text{ex}}$: $\Delta E_{\text{esc}} = 3.28 J_{\text{ex}}$, once again for all three types of skyrmions. It is worthy to note that in the present configuration, this mechanism possesses four equivalent realizations (one at each side of the square), which makes it more likely by a factor of four in the rate prefactor.

\paragraph{Curved boundary.}Fig. \ref{fig:curved_boundary_mechanism} shows the escape of a Néel skyrmion through a curved boundary. The nature of the discrete lattice means that the boundary exhibits a staircase effect. The escape process is similar to that of the flat boundary configuration, but the activation energy is found to be slightly higher: $\Delta E_{\text{curv}} = 3.60 J_{\text{ex}}$. Once again, four equivalent processes exist. The energy profile appears very similar to that of Fig. \ref{fig:Etot_boundary} (see the Supplemental Material \cite{sm}).

\section{Rate prefactor}\label{sec:prefactor}
\paragraph*{}As discussed in the introduction, estimating activation barriers is not sufficient in order to obtain the average lifetime of magnetic structures. Knowledge of a rate prefactor is also recquired. In what follows, we present the basis for the theory behind the computation of the different terms of that prefactor. These are the ratio of curvatures of the energy at $A$ and $S$ that we obtain from the diagonalization of the Hessian matrix, and a dynamical contribution that comes from the deterministic equations of motion linearized about the saddle point.
 
\subsection{Ratio of energy curvatures from the Hessian matrix.} We consider an assembly of $N$ magnetic spins of constant amplitudes and described by a set of $2N$ variables that we write in the form of a row vector $\boldsymbol{\eta} = (\eta_1 \dots \eta_{2N}).$ 
One important assumption in Langer's theory is that the energy of the system close to the saddle point and the metastable minimum can be approximated as a Taylor series truncated to second order,

\begin{equation}\label{eq:taylorseries}
E(\boldsymbol{\eta}) \sim  E^{0} (\boldsymbol{\tilde{\eta}}) + \frac{1}{2} \big( \boldsymbol{\eta}-\boldsymbol{\tilde{\eta}}\big) H_{ \boldsymbol{\tilde{\eta}}}
\big( \boldsymbol{\eta}-\boldsymbol{\tilde{\eta}}\big)^T,
\end{equation}
where $\boldsymbol{\tilde{\eta}} = (\tilde{\eta}_1 \dots \tilde{\eta}_{2N}) $ are the coordinates of a local extremum (the local minimum $A$ or saddle point $S$). At the extrema, 
\begin{equation}
\dfrac{\partial E}{\partial \boldsymbol{\eta}}_{\big| \boldsymbol{\tilde{\eta}}} = 0,
\end{equation}
in which the notation $_{\big| \boldsymbol{\tilde{\eta}}}$ means the expression is evaluated at $\boldsymbol{\tilde{\eta}}$ and
\begin{equation}\label{eq:hessian}
H_{ \boldsymbol{\tilde{\eta}}} = 
\begin{pmatrix}
    \dfrac{\partial^2 E}{\partial \eta_1^2}_{\big| \boldsymbol{\tilde{\eta}}} & \dots &  \dfrac{\partial^2 E}{\partial \eta_1 \partial \eta_{2N}}_{\big| \boldsymbol{\tilde{\eta}}} \\
         \vdots       &   & \vdots \\
   \dfrac{\partial^2 E}{\partial \eta_{2N} \partial \eta_1}_{\big| \boldsymbol{\tilde{\eta}}}        & \dots & \dfrac{\partial^2 E}{\partial \eta_{2N}^2}_{\big| \boldsymbol{\tilde{\eta}}} \\
\end{pmatrix}
\end{equation}
is the energy Hessian evaluated at $\boldsymbol{\tilde{\eta}}$  which contains the second derivatives of the energy. It is symmetric and real, and therefore Hermitian by construction. Details concerning our implementation of the Hessian in spherical coordinates on the unit sphere $(1,\theta,\phi)$ are given in appendix \ref{app:hessian}. 
The $\{\lambda_i$\} are the eigenvalues of the Hessian and correspond to the $2N$ curvatures of the energy surface in normal mode space. A positive (negative) curvature corresponds to a mode of stable (unstable) fluctuations. A zero-curvature corresponds to a Goldstone mode of zero energy fluctuation and is associated with a continuous unbroken global symmetry \cite{langer,braun1994kramers}. The corresponding Gaussian integral becomes $\int \diff a_i$
evaluated over all possible values of the associated eigenfunction coordinate $a_i$ and needs to be handled separately. It also yields an additional $\sqrt{2\pi k_B T}$ factor in the ratio of eigenfrequencies and consequently makes the rate prefactor in Eq. (\ref{eq:escaperate}) temperature-dependent. For first order saddle points, all curvatures at $A$ and $S$ are either positive or zero-curvatures, aside from a single negative curvature at the top of the barrier. This unstable mode is the one that will eventually allow the system to escape over the barrier and to the lower energy minimum.

\paragraph*{}If there are no zero-curvatures, the factor $\Omega_0$ in Eq. (\ref{eq:prefactor}) is then obtained from the squareroot of the ratio of determinants of the Hessian at $A$ and $S$,
\begin{equation}\label{eq:omega0_nozero}
\Omega_0 = \sqrt{\frac{ \det  H^{A}}{| \det H^{S}|}} = \sqrt{\frac{\prod_i\lambda_i^A} {\prod_j|\lambda_j^S|}}.
\end{equation}
Examples on how the theory is extended to include Goldstone modes can be found in previous works by Braun \cite{braun1994kramers} and Loxley \cite{loxley}. 

\subsection{Dynamical prefactor $\lambda_+$.} The dynamical prefactor takes into account the dynamics of the system at the saddle point and is derived from the set of $N$ deterministic Landau-Lifshitz-Gilbert (LLG) equations linearized at the saddle point. We obtain
\begin{equation}
\frac{\diff}{\diff t}
\begin{pmatrix}
\eta_1 - \tilde{\eta}_1 \\
\vdots \\
\eta_{2N} - \tilde{\eta}_{2N}
\end{pmatrix}
=
\mathcal{T}
\begin{pmatrix}
\eta_1 - \tilde{\eta}_1 \\
\vdots \\
\eta_{2N} - \tilde{\eta}_{2N}
\end{pmatrix}
\end{equation}
in which $\mathcal{T}$ is the transition matrix of LLG. Details of the derivation and the analytical expression of $\mathcal{T}$ in spherical coordinates are given in appendix \ref{app:LLG}. Similarly to the Hessian, it is a $2N \times 2N$ matrix but possesses $2N-1$ negative eigenvalues associated with the stable modes, and a single positive eigenvalue $\lambda_+$, which gives the growth rate the dynamically unstable deviation at the saddle \cite{coffey}. We can note that the transition matrix is not symmetric and can in principle admit complex eigenvalues and eigenvectors.


\section{Results}\label{sec:results}

\begin{table*}
\centering
\caption{Terms of the rate prefactor and total annihilation rate at $T$=300K for all mechanisms. The size of the simulated domain is chosen as not to impact the transition rate. $\Omega_{0,\text{int}}$ gives the contribution of internal modes to the prefactor and $\Omega_{0,\text{tot}}$ gives the total contribution of all modes. $\Gamma_0$ and $\Gamma$(300K) are calculated using $\Omega_{0,\text{tot}}$. The attempt frequency is multiplied by four to account for all equivalent realizations of boundary escape mechanisms. $\Gamma$(300K) is calculated for $J_{\text{ex}}$ = 1.6$\times 10^{20}$ J.\label{tab:transition_rates}}
\subcaption{Collapse.}
\begin{ruledtabular}
\begin{tabular}{cccccccc}
Mechanism & $\Delta E$ ($J_{\text{ex}}$) & $\Omega_{0,\text{int}}$  & $\Omega_{0,\text{tot}}$ & $\Omega_{0,\text{int}} / \Omega_{0,\text{tot}}$ & $\lambda_+$ (GHz) & $ \Gamma_0$ (MHz) & $\Gamma$(300K) (kHz)   \\
\hline
single sk & 2.83  & 0.0015 & 3.51 $\times 10^{-5}$ & 43 & 1200.47 &  6.70 & 0.12  \\ 
two sk. & 2.82  &  0.0009 & 2.32  $\times 10^{-5}$ & 42 & 1200.23 & 4.43  & 0.08 \\
defect & 0.38 & 0.0214 & 1.16$\times 10^{-3}$ & 18 & 145.41 & 26.90 & 6190.40
  \\
\end{tabular}
\bigskip
\subcaption{Escape.}
\begin{tabular}{cccccccc}
Mechanism & $\Delta E$ ($J_{\text{ex}}$) & $\Omega_{0,\text{int}}$  & $\Omega_{0,\text{tot}}$ & $\Omega_{0,\text{int}} / \Omega_{0,\text{tot}}$ & $\lambda_+$ (GHz) & $ \Gamma_0$ (MHz) & $\Gamma$(300K) (kHz)   \\
\hline
flat bound. & 3.28  & 0.0349 & 1.24  $\times 10^{-2}$ & 2.8 & 522.94 & 4144.6 &  13.00   \\
curved bound. &  3.60 & 0.0198 & 4.48 $\times 10^{-3}$  & 4.4 &  501.45 &      1428.84
 & 1.29 \\
\end{tabular}
\end{ruledtabular}
\end{table*}

\begin{figure}[hbtp]
\centering
		\includegraphics[width=.9\textwidth]{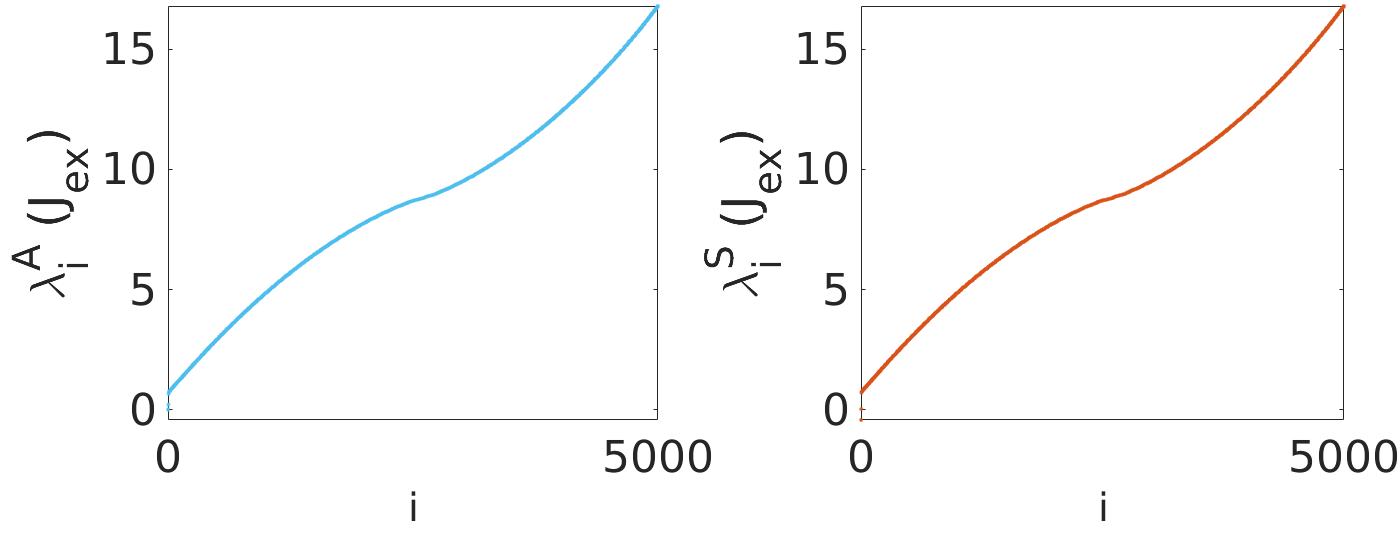}
		\caption{All energy curvatures at $A$ (in blue) and  $S$ (in red) ordered by increasing amplitudes for the case of the collapse of a single skyrmion. The other mechanisms exhibit the same profiles which we interpreted as the dispersion of spin-wave excitations, with the exception of the first few eigenvalues. These are found below the main curve and are shown on Fig. \ref{fig:curva_all}a. }
		\label{fig:curvatures_all}
\end{figure}

\begin{figure*}[hbtp]
\centering
	\begin{subfigure}[t]{.33\textwidth}
		\adjincludegraphics[width=1\textwidth,trim={0 0 0 {.26\height}},clip]{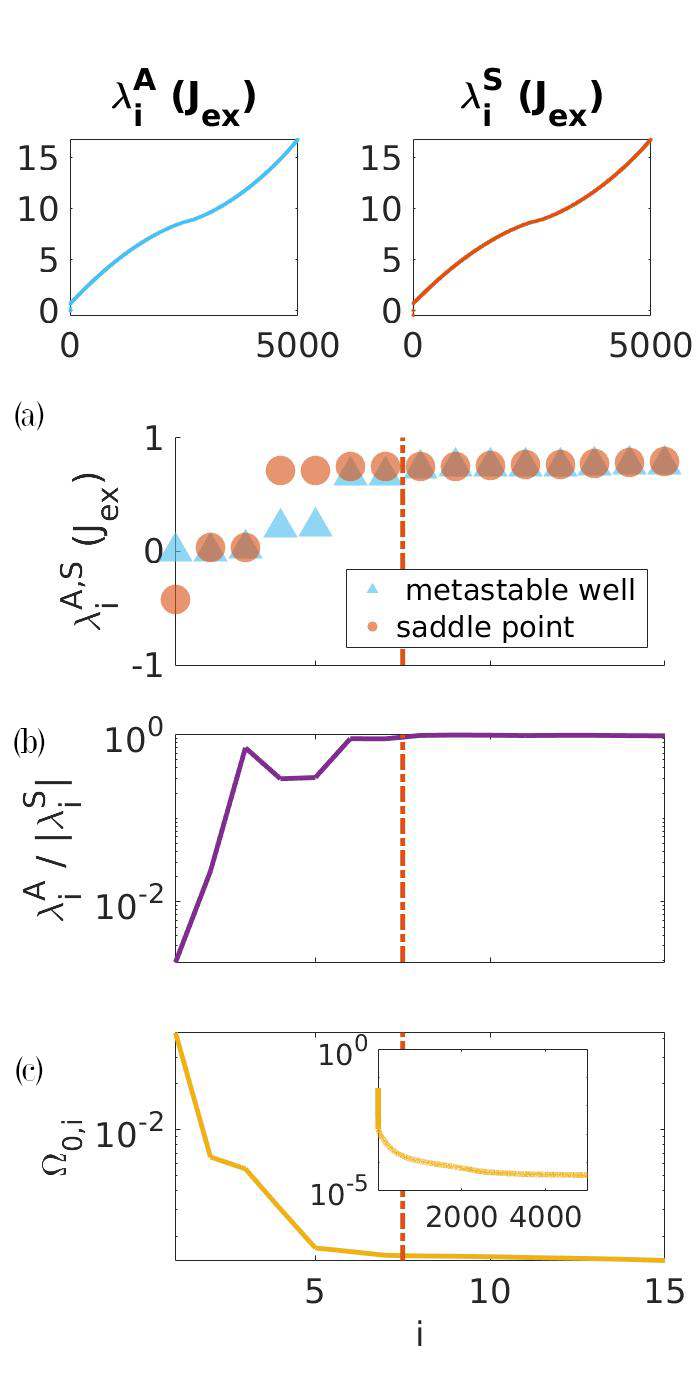}
\caption*{Collapse of a single skyrmion.}
	\label{fig:curva_iso}
	\end{subfigure}\hfill	
	\begin{subfigure}[t]{.33\textwidth}
			\adjincludegraphics[width=1\textwidth,trim={0 0 0 {.26\height}},clip]{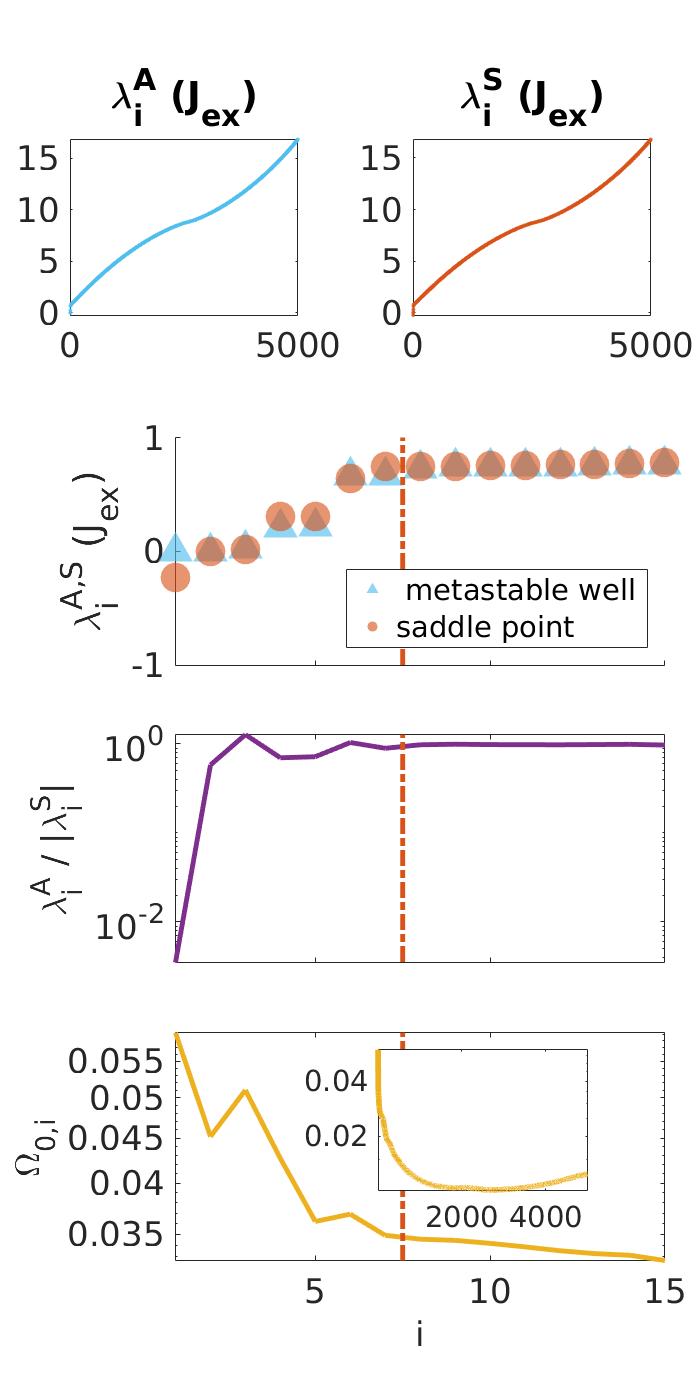}
		\caption*{Escape through a flat boundary.\label{fig:curva_boundary}}	
	\end{subfigure}\hfill	
	\begin{subfigure}[t]{.33\textwidth}
		\adjincludegraphics[width=1\textwidth,trim={0 0 0 {.26\height}},clip]{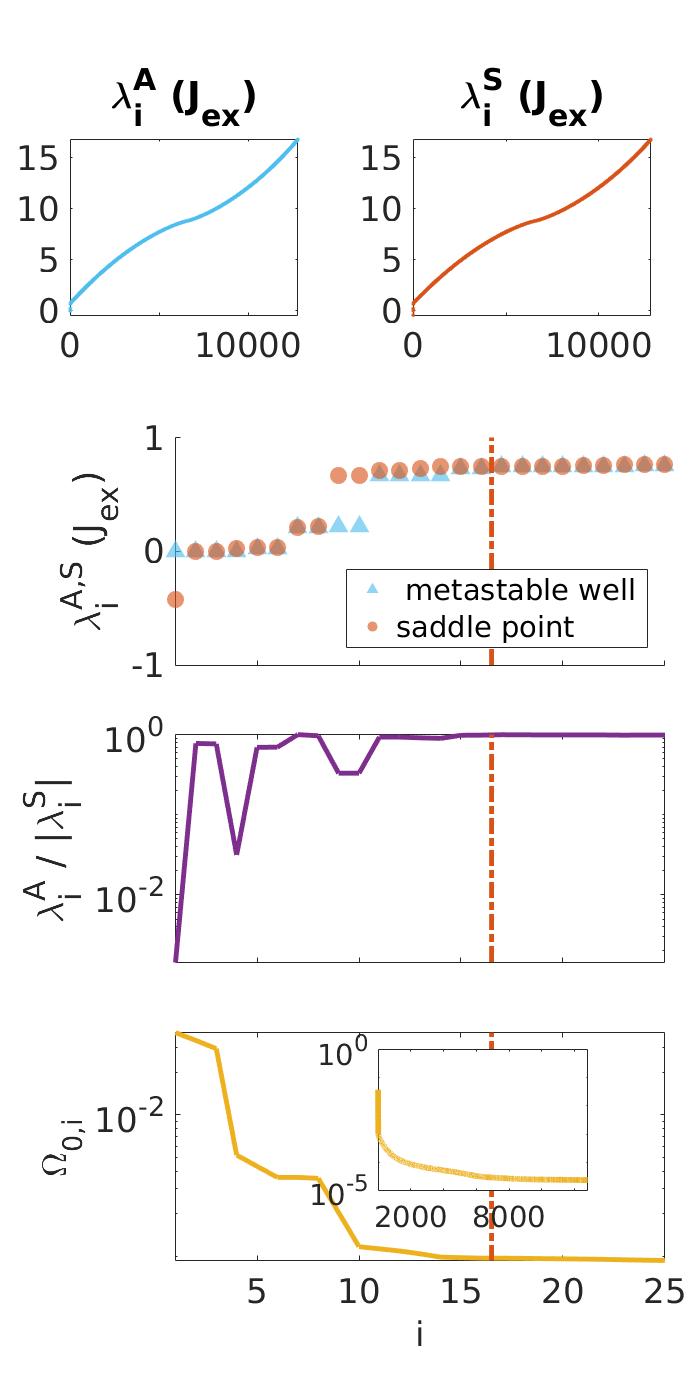}
		\caption*{Collapse in the presence of another skyrmion.}
		\label{fig:curva_2sk}
	\end{subfigure}
\caption{For the first few eigenvalues of each annihilation mechanism, we show the following: (a) Eigenvalues at $A$ and $S$. (b) Ratio of individual eigenvalues in semilog scale. (c) $\Omega_{0,i}$ in semilog scale. The inset figure shows all eigenvalues. The red dotted line marks the separation between localized and collective eigenmodes. The $x$-axis is the same for all subfigures.}
\label{fig:curva_all}
\end{figure*}

\paragraph*{}Once the saddle point is accurately identified along a path, the corresponding rate prefactor can be calculated. 
All terms entering in the calculation of activation rates are summarized in Table  \ref{tab:transition_rates} for each mechanism. Collapse and escape through a flat boundary were studied for the Néel and Bloch skyrmions, as well as the antiskyrmion, and yield the same results. In the cases of the escape through a curved boundary and the collapse on a defect, we studied only the Néel skyrmion, but it seems reasonable to assume the following results also hold for other types of topological defects. A possible issue of non-negligeable numerical rounding errors affecting the accuracy of the ratio of eigenvalues was previously mentioned in Ref. \onlinecite{seuss}. In  Table \ref{tab:lattice_sizes} of Appendix \ref{app:simulations}, we gather results of simulations performed for different lattice sizes, and show that as long as the skyrmion is not  constrained by the boundary, and as long as Goldstone modes do not arise, the size of the lattice does not significantly affect the computed attempt frequencies. In each case we consider, the curvatures are ordered by increasing amplitude with corresponding index $i$ and plotted on Fig. \ref{fig:curvatures_all} for all $i$ and on Fig. \ref{fig:curva_all}a for the first 15 or 25 values of $i$. Fig. \ref{fig:curva_all}b shows the ratio of individual curvatures plotted in semilog scale. The value of $\Omega_{0,i} = \sqrt{\prod_{j=1}^i \lambda^A_j / |\lambda^S_j|}$ is plotted in a similar fashion on Fig. \ref{fig:curva_all}c where all the curvatures are shown in the inset figure. The same graphs for the collapse in the presence of a defect and escape through a curved boundary can be found in the Supplemental Material \cite{sm}, and the following conclusions also apply. We note that for all mechanisms, the ratio of curvatures only shows significant variations for small $i$ and weak variations for $i \gg 1$.  Consequently, the value of $\Omega_{0,i}$ shows a strong $i$-dependence at small $i$ and a weak dependence for larger $i$. More specifically, for collapse mechanisms, $\Omega_{0,i}$ shows a strong $i$-dependence for small $i$, a medium dependence for intermediate $i$ and a weak dependence for large $i$. In the case of escape mechanisms, the $i$-dependence has a different profile. It appears stronger for intermediate indices and, after first decreasing, goes up again in the domain of highest curvatures.


\begin{figure*}
\centering
	\begin{subfigure}[t]{.33\textwidth}
		\includegraphics[width=1\textwidth]{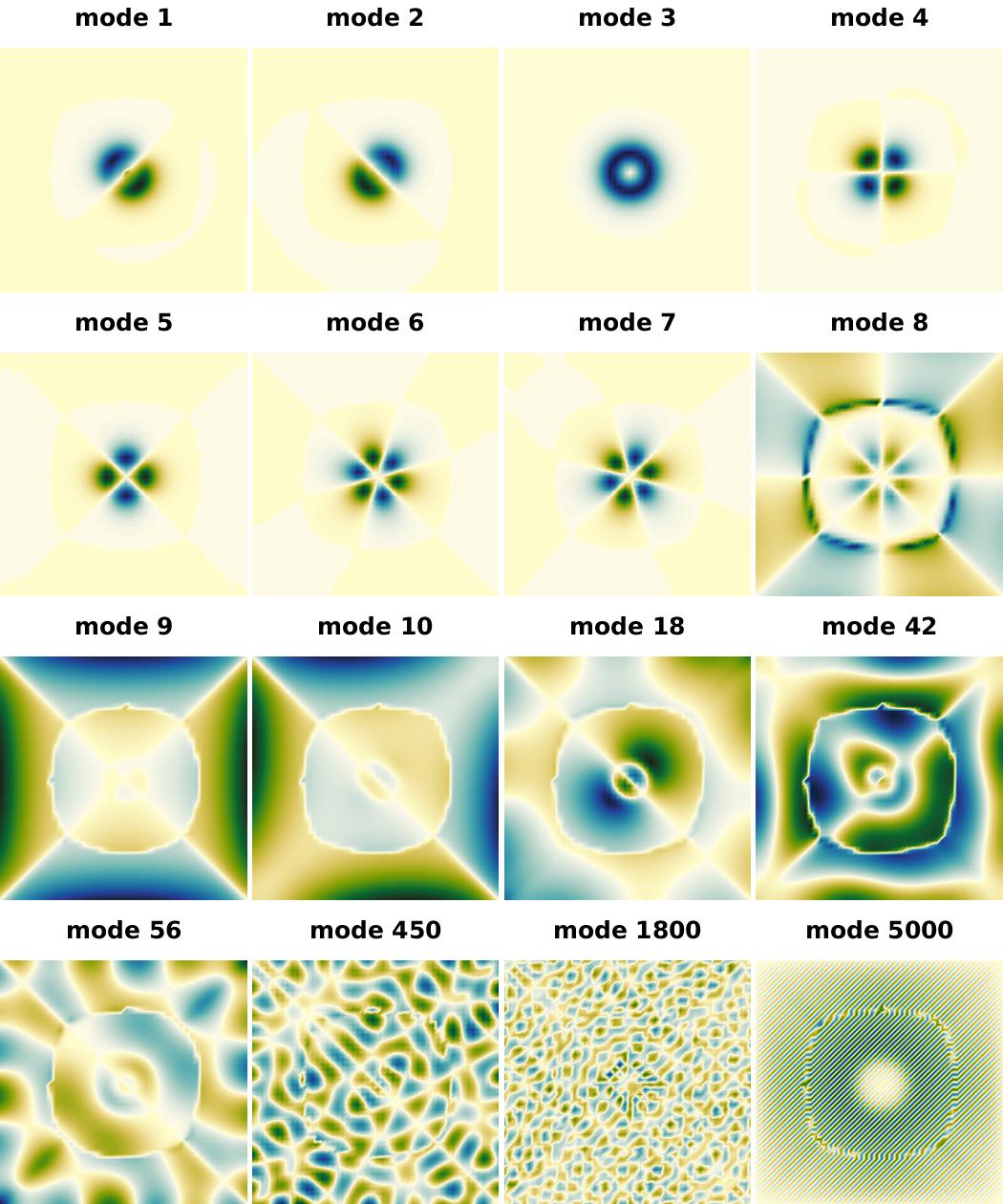}
\caption{}
	\label{fig:meta_modes_iso}
	\end{subfigure}\hfill
	\begin{subfigure}[t]{.33\textwidth}
		\includegraphics[width=1\textwidth]{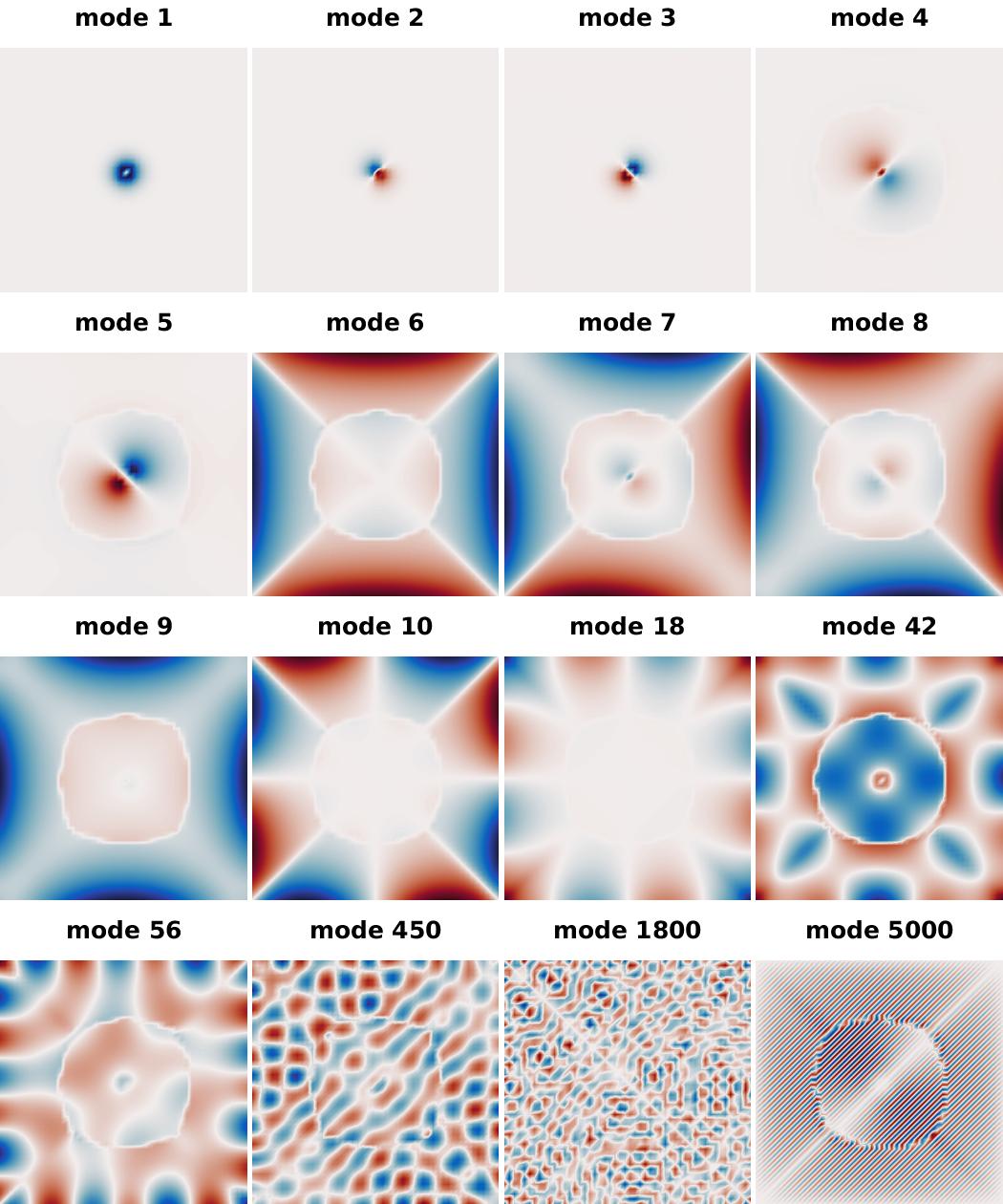}
		\caption{}
	\label{fig:SP_modes_iso}
	\end{subfigure}\hfill
			\begin{subfigure}[t]{.33\textwidth}
		\includegraphics[width=1\textwidth]{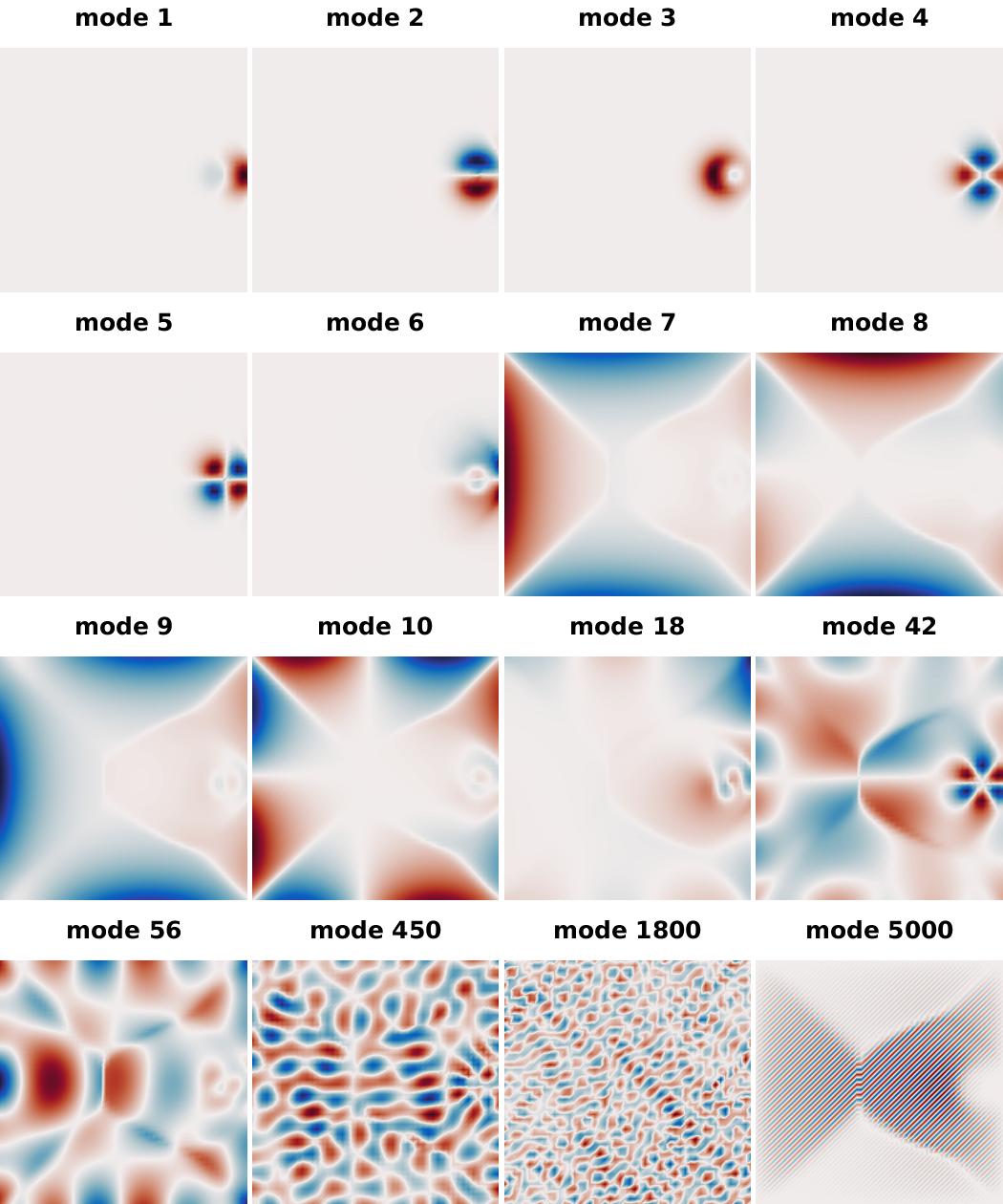}
\caption{}
	\label{fig:SP_modes_bound}
	\end{subfigure}\hfill
	\caption{Eigenmodes associated to the $\theta$ variable for a single skyrmion in a  simulated system of $50 \times 50$ spins. The blue and green color scheme is associated with metastable states and the blue and red one with saddle points. Negative amplitudes are plotted in blue and positive ones in green/red (color online). The range of the color map is adjusted on each plot so that zero-amplitude fluctuations coincide with white. Modes are designated via the $i$-index of the corresponding ordered eigenfrequencies of Fig. \ref{fig:curva_all}. (a) Metastable single-skyrmion state  with localized modes $i = {1 \dots 7}$ and collective modes $i > 7$. (b) Saddle point of the collapse with localized modes $i = {1 \dots 5}$ and collective modes $i > 5$. (c) Saddle point of the escape with localized modes $i = {1 \dots 6}$ and collective modes $i > 6$. \label{fig:modes_1sk}}			
\end{figure*}
	
	
	\begin{figure}
\centering
\begin{subfigure}[t]{.9\textwidth}		
		\includegraphics[width=1\textwidth]{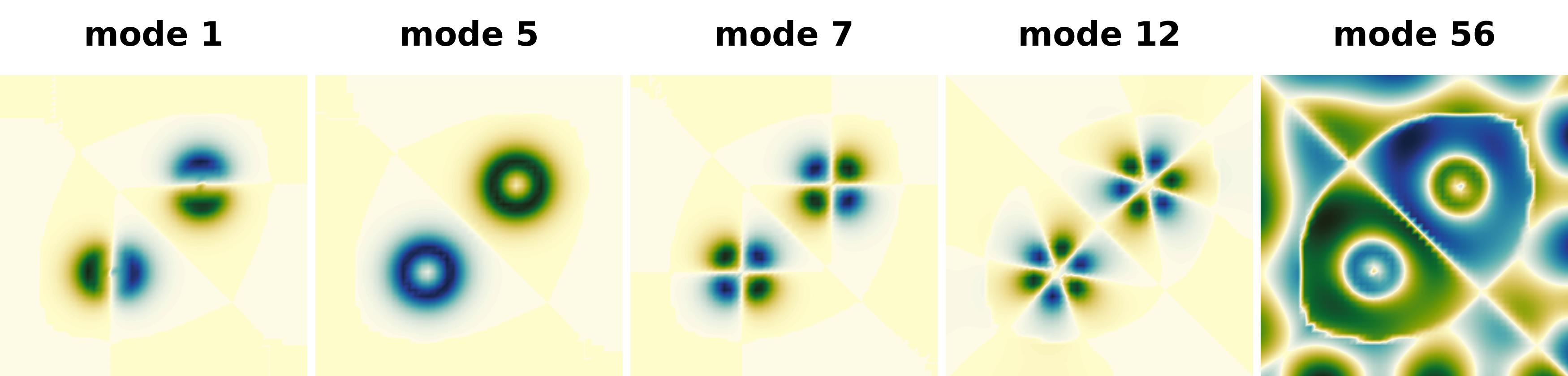}

		\caption{}
		\label{fig:meta_modes_2sk}
	\end{subfigure}
		
	\begin{subfigure}[t]{.9\textwidth}		
		\includegraphics[width=1\textwidth]{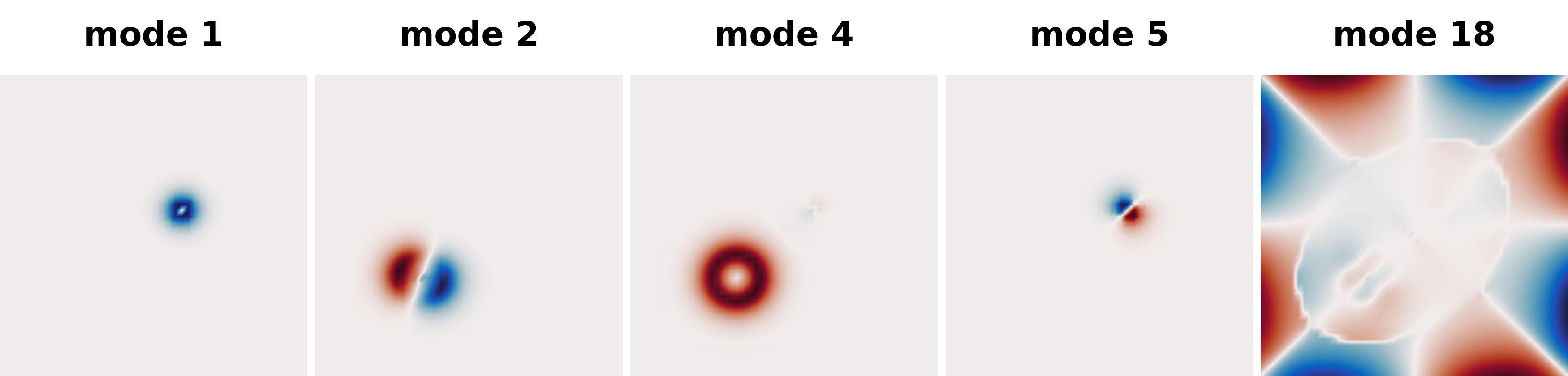}
		\caption{}
		\label{fig:SP_modes_2sk}
	\end{subfigure}
	
	\begin{subfigure}[t]{.45\textwidth}
		\includegraphics[width=1\textwidth]{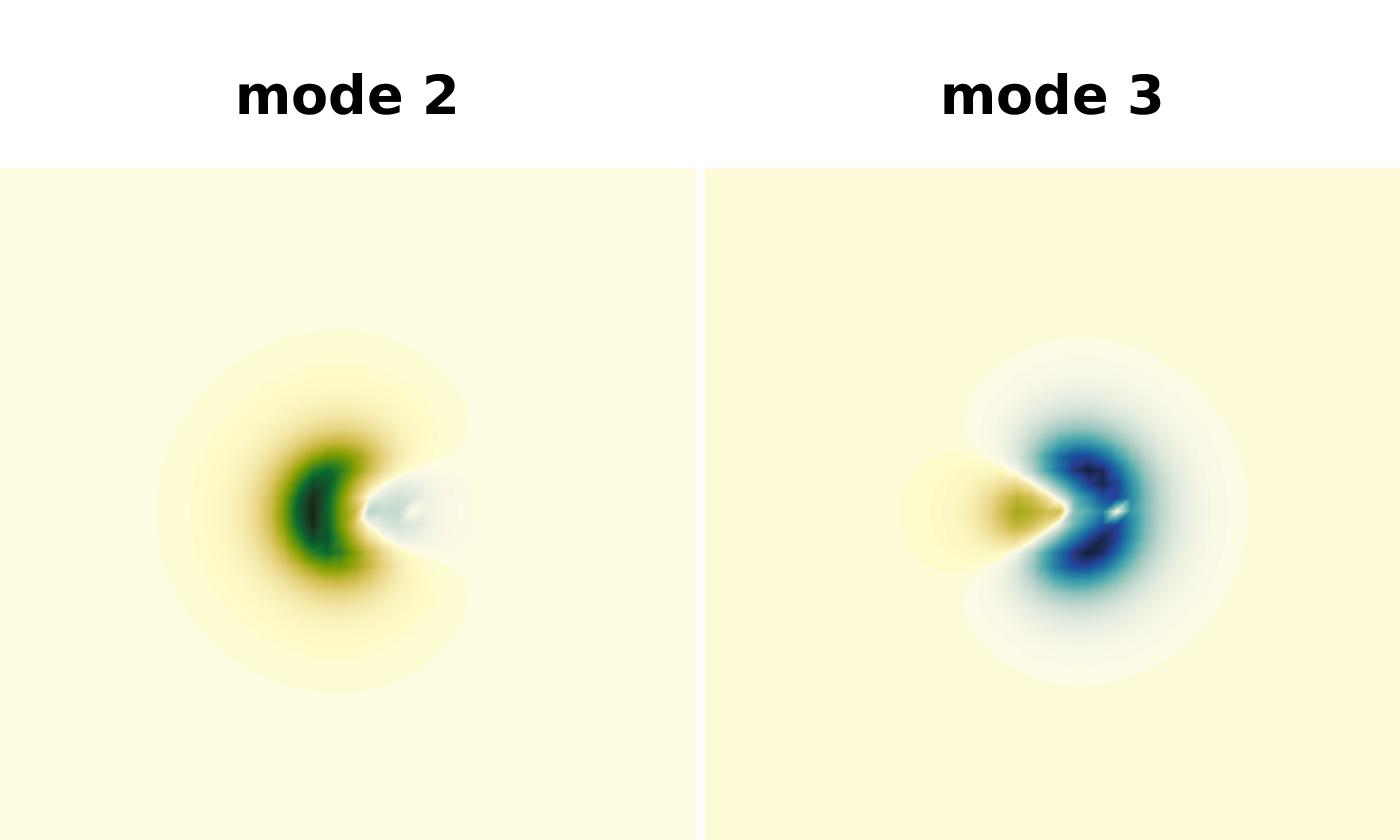}
\caption{}
	\label{fig:meta_modes_def}
	\end{subfigure}\hfill
	\begin{subfigure}[t]{.45\textwidth}
		\includegraphics[width=1\textwidth]{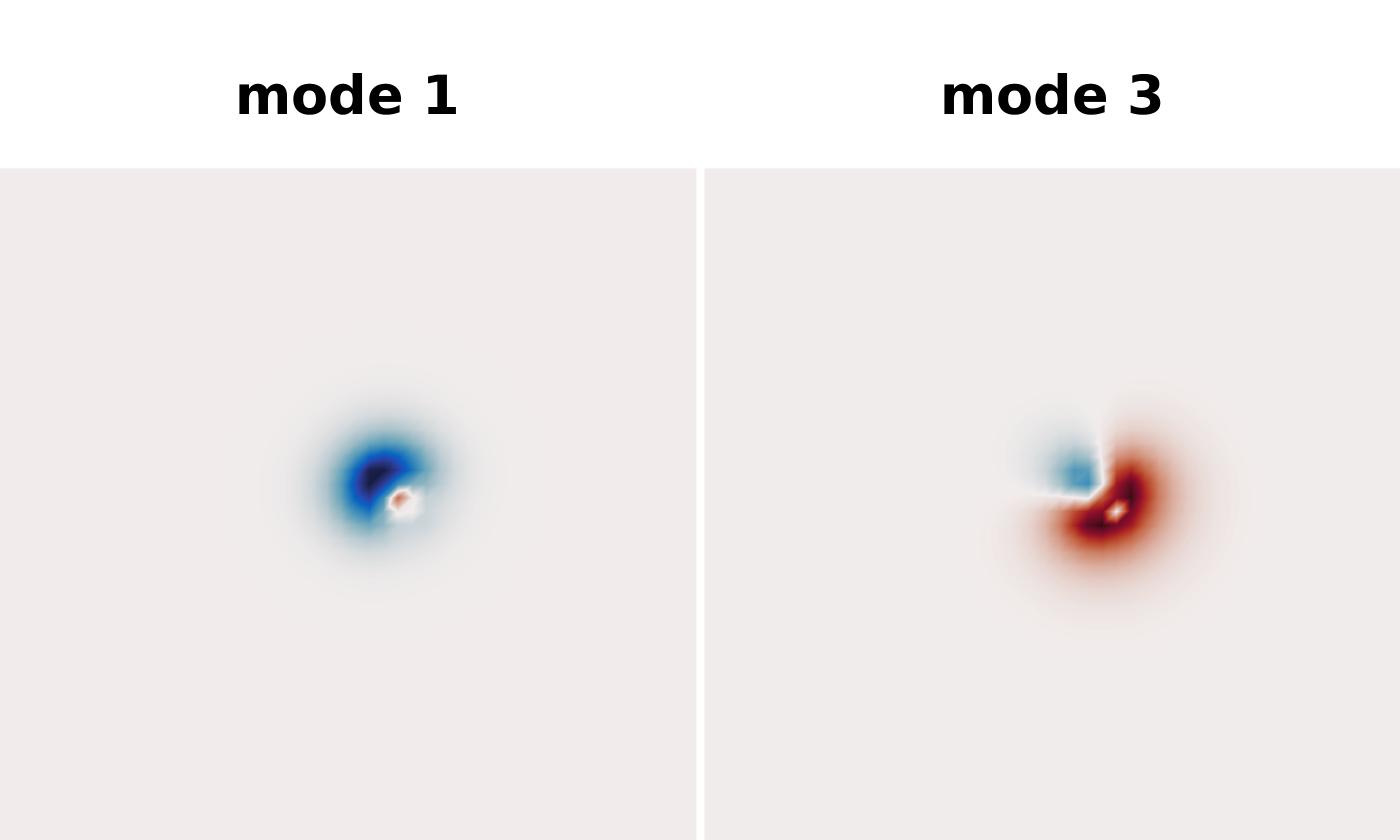}
\caption{}
	\label{fig:SP_modes_def}
	\end{subfigure}\hfill
	
	\begin{subfigure}[t]{.9\textwidth}
		\includegraphics[width=1\textwidth]{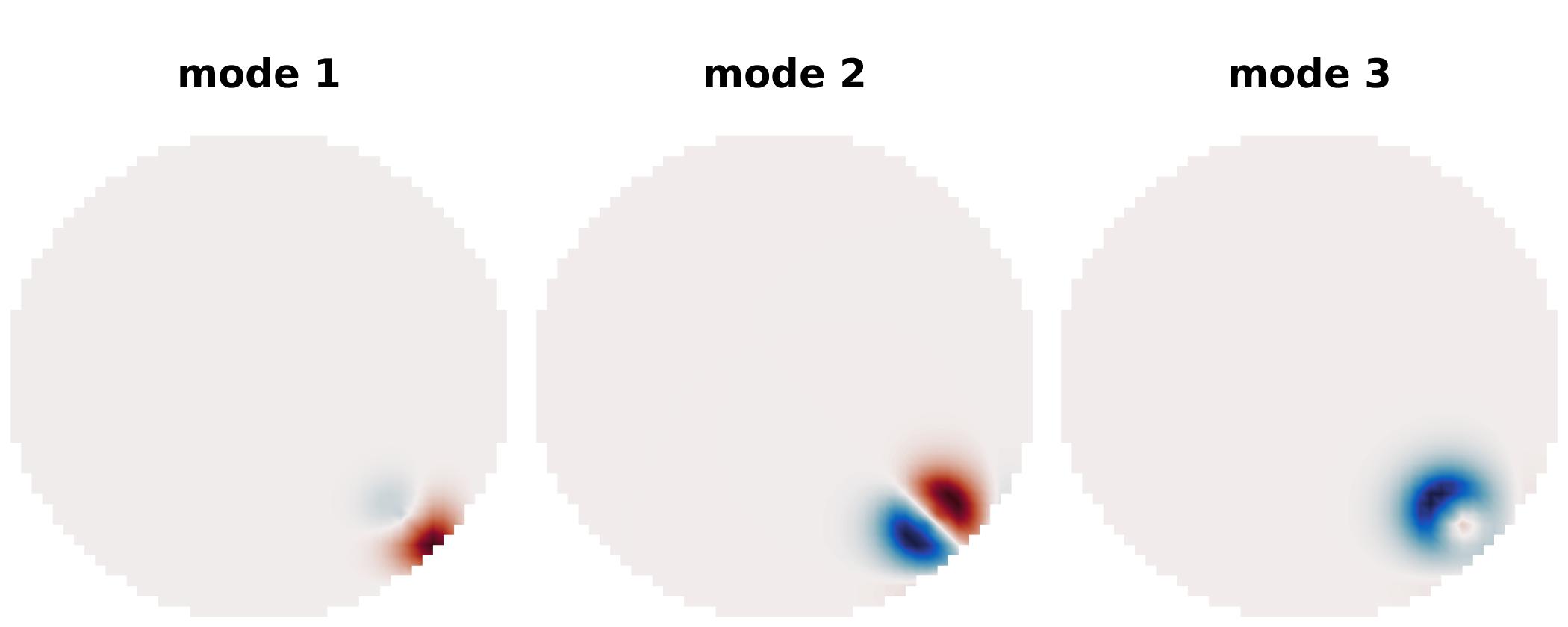}
\caption{}
	\label{fig:SP_modes_cb}
	\end{subfigure}
	\caption{Eigenmodes associated to the $\theta$ variable. The color code is the same as in Fig. \ref{fig:modes_1sk}. Additional mode profiles can be found in the Supplemental Material \cite{sm} for each of the following mechanisms. (a, b) Internal and collective modes of two coupled skyrmions on a $50 \times 50$ lattice at $A$ and $S$. (c, d) Modified internal modes for the collapse on a defect at $A$ and $S$. (e) Modified internal modes for the escape through a curved boundary at S. \label{fig:modes_modif}}	
\end{figure}

\subsection{The thermal role of internal eigenmodes.} In order to obtain the spatial profiles of the eigenmodes, we expand small fluctuations of the spin orientations in the eigenbasis,
\begin{equation}\label{eq:eigenbasis}
\eta_i - \tilde{\eta}_i = a_i  x_i
\end{equation}
where $\{x_i\}, i=1\dots 2N$ is a set of orthonormal eigenvectors forming a complete basis in the space of configurations. The relative amplitudes of small fluctuations about the saddle point and the metastable state for each mode $i$ are thus contained in the components of the corresponding eigenvector $x_i$. This allows us to plot the spatial profiles for the $\theta$-eigenmodes at $A$ and $S$ on Fig. \ref{fig:modes_1sk} and \ref{fig:modes_modif}.  Additional mode profiles can be found in the Supplemental Material \cite{sm}. The $\phi$-profiles exhibit similar behaviour and do not provide any further information for the following analysis. In recent years, skyrmions' eigenmodes have been extensively investigated  \cite{makhfudz2012inertia,lin2014internal,schutte2014magnon,iwasaki2014theory,buijnsters2014zero,schroeter2015scattering,zhang2017eigenmodes,mruczkiewicz2017spin,guslienko2017gyrotropic,garst2017collective,kravchuk2018spin}.
Localized modes were reported to exist below the magnon continuum, with excitations such as translational motion, uniform breathing mode, elliptic and triangular distortions, etc. Classifications based on the azimuthal number $m$ were proposed, which is linked to the number of nodes $2\lvert m\rvert $ encountered when going around the skyrmion center along the azimuthal angle. In this work, we use the $i$-index of the ordered eigenfrequencies (Fig. \ref{fig:curva_all}) to classify the modes.  While some relation to spin waves calculated in previous work could be made, our classification scheme is useful because it allows comparison of eigenvalues and eigenfunctions of the fluctuations around both the metastable state and the saddle point. 
On Fig.  \ref{fig:modes_1sk},  and in agreement with previous studies, the first clear observation is that for all cases, the lowest frequency eigenmodes are localized internal skyrmion modes. The rest of the modes are collective modes extended to the entire lattice and can be thought of as amplitudes corresponding to spin-wave (SW) excitations. The numerical values of the eigenfrequencies associated to the internal modes can be found in the Supplemental Material \cite{sm} for the metastable single skyrmion state and the saddle points of the escape and collapse mechanisms. 
  
The first two modes of the metastable skyrmion state are translational modes ($m=\pm 1$) (Fig. \ref{fig:meta_modes_iso}). These are not zero-modes  because there exists no exact translation symmetry here, however the corresponding eigenvalues are close to zero (see Fig. \ref{fig:curva_all}a and Table I. of the Supplemental Material \cite{sm}). Mode 3 is the uniform breathing  mode ($m =0)$ and is a low frequency mode slightly above modes 1 and 2 on Fig. \ref{fig:curva_all}a. The next two modes directly above them correspond to elliptic distortions of the skyrmion shape $(m = \pm 2)$ and the final two local modes are triangular distortions $(m = \pm 3)$. Therefore, there are seven local states of the metastable well. The following modes are part of the magnon continuum. At the saddle point of the collapse mechanism (Fig. \ref{fig:SP_modes_iso}), the unstable mode becomes the uniform breathing mode. Mode 2 and 3 are translational modes. In total, five local states exist at the saddle point. 
 
At the boundary (Fig. \ref{fig:SP_modes_bound}), the unstable skyrmion exhibits similar modes to that of its metastable counterpart, but they appear distorted by the presence of the edge. Six local modes are found. The unstable mode (mode 1) is a mode of translation towards the boundary and is the mode that enables the escape of the skyrmion. The edge lifts the degeneracy of the translational modes: mode 2 is a mode of translation parallel to the boundary, and the associated eigenfrequency is close to zero (Fig. \ref{fig:curva_all}a) as this displacement costs little energy. 
Whether the boundary is flat or curved (Fig. \ref{fig:SP_modes_cb}), the same number of internal modes exist at $A$ and $S$ and they appear similar. There are of course fewer collective modes associated to the circular sample as there are fewer magnetic sites and therefore fewer degrees of freedom.

 Fig. \ref{fig:meta_modes_def} and \ref{fig:SP_modes_def} display examples of how the presence of a defect modifies some of the internal modes. 
All the internal modes in that case can be found in the Supplemental Material \cite{sm}. In particular, the unstable breathing mode at the saddle point is significantly affected by the presence of the defect (Fig. \ref{fig:SP_modes_def}). The defect also lowers the total number of internal modes, as six internal modes survive at $A$ and four at $S$. 
  In the system of two metastable coupled skyrmions, the internal modes are the same as that of a single skyrmion and both skyrmions are excited simultaneously (Fig. \ref{fig:meta_modes_2sk}). This is no longer the case at the transition state (Fig. \ref{fig:SP_modes_2sk}) where all the amplitude is localized to only one of the two skyrmions in each of the internal modes. Note that Fig. \ref{fig:meta_modes_2sk} and \ref{fig:SP_modes_2sk} show modes on the $50 \times 50$ lattice whereas calculations were carried out for the the $80 \times 80$ lattice. 
 
\paragraph*{}On Fig. \ref{fig:curva_all}a, the internal modes match the modes appearing below the magnon continuum of Fig. \ref{fig:curvatures_all}. The separation between localized and collective modes is shown by a dashed line on Fig. \ref{fig:curva_all} and also coincides with the transition between strong and low $i$-dependence of $\Omega_{0,i}$ (Fig. \ref{fig:curva_all}c). The contribution of the internal modes to the prefactor is given in Table \ref{tab:transition_rates} by $\Omega_{0,\text{int}} $ while the complete contribution of all the modes corresponds to $\Omega_{0,\text{tot}}$. For the collape mechanisms, the values differ by a factor of $\sim 40$ between them, or 18 when a defect is present, whereas in the case of the escape through the boundary, it is only a factor of three (flat boundary) or four (curved boundary). In other words, the relative contribution of internal modes is higher for collapse processes compared to the case of an escape through a boundary.

\paragraph*{}From all the above observations, we conclude that internal modes play the most significant role in the thermally activated annihilation of a skyrmion. Each of the spin-wave modes brings a weak contribution but because there are many more SW modes than there are internal modes, their contribution to the attempt frequency cannot be neglected. In the high frequency domain, the wavelength of the SW modes is much smaller than the radius of the skyrmion, therefore the contribution of the highest frequency modes is smaller (see the last couple of modes on each subfigure of Fig. \ref{fig:modes_1sk} and in the Supplemental Material \cite{sm}). This seems not to be true in the boundary annihilation where high frequency modes appear to contribute more than the intermediate ones. One possible explanation is that the coupling to the boundary in the saddle configuration means the impact of spin-waves on the skyrmion is more important. Additionally, the contribution of the internal modes to the attempt frequency is higher in collapse processes compared to annihilation at the edge. It is however reduced by half if a defect is present. This hints at the fact that the relative contribution of internal and SW modes is strongly linked to the nature of the annihilation and the geometry of the transition state.

\subsection{Broken symmetries.}Our second observation concerns the symmetries: the eigenmodes at the saddle point tend to display broken symmetries compared to the metastable states. In the single metastable skyrmion case (Fig. \ref{fig:meta_modes_iso}), the internal modes in particular possess symmetries of types two-fold, four-fold, six-fold, and radial. At the saddle point of the  collapse (Fig. \ref{fig:SP_modes_iso}), four-fold and six-fold symmetries are gone. The breathing mode (mode 1), which is unstable at the SP, displays a broken radial symmetry with a distorted center. That broken symmetry pattern around the center is also visible in many higher frequency modes. As for the saddle configuration at the boundary (Fig. \ref{fig:SP_modes_bound} and \ref{fig:SP_modes_cb}), symmetries are broken by the edge. Lastly,  in the case of the two skyrmions (Fig. \ref{fig:SP_modes_2sk}), the symmetry breaking at the saddle is striking as each internal mode involves only one skyrmion, in contrast to the metastable eigenmodes.

\begin{figure}[hbtp]
\centering
		\includegraphics[width=.8\textwidth]{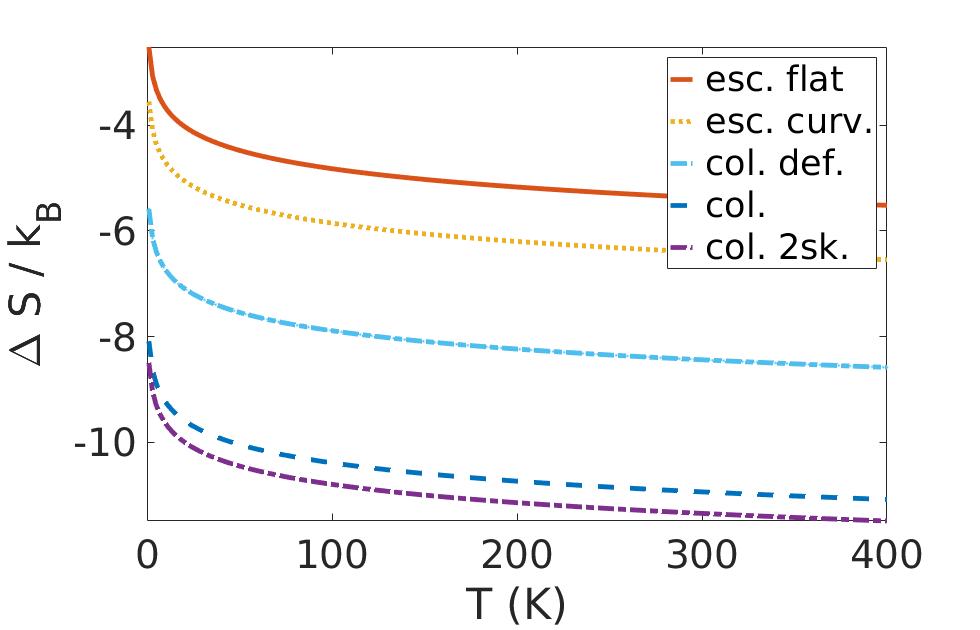}
		\caption{Calculated change in configurational entropy induced when the system goes to the saddle point $\Delta S / k_B = \dfrac{S_S - S_A}{k_B}$ as defined in Eq.  (\ref{eq:config_entropy}) over a broad range of temperatures for all mechanisms. The highest entropic barrier corresponds to the most negative $\Delta S$, \textit{ie} the collapse involving a second skyrmion.}
		\label{fig:deltaS}
\end{figure}

\subsection{Entropic contribution and skyrmion stability. }Excluding the collapse on a defect, for which the energy barrier is lowered significantly, we find that escape processes, whether through a flat or curved boundary, are the most probable mechanisms, even though they paradoxically involve both the highest activation energies and the lowest characteristic times in the dynamics about the saddle. Therefore, and as has been peviously discussed \cite{wild2017entropy,hagemeister2015stability}, activation energies alone do not allow the lifetime of skyrmions to be predicted. Characteristic times at the transition state were found to lie in the GHz-THz regime, which remains in the range of typically assumed values for estimates of the prefactor in magnetic spin systems. Yet, a large difference is observed here due to the contribution of the ratio of curvatures $\Omega_0$, which significantly lowers the attempt frequency. To interpret this result and the meaning behind a low value of $\Omega_0$, we examine Eq. (\ref{eq:omega0_nozero}). The product of curvatures evaluated at an extremum in the energy surface $\big(\prod_i \lambda^{A,S}_i\big)^{-1}$ can be seen as a measure of the total volume of configuration space ($\eta$-space) accessible to thermal fluctuations in that particular state. The ratio of eigenvalues thus corresponds to the change in that volume induced by the transition from $A$ to $S$. In other words, it characterizes the relative volume of the saddle point region. A low value of $\Omega_0$ is associated with a large volume of the metastable skyrmion well and/or a narrow saddle region in $\eta$-space (eg: Fig. \ref{fig:3d_landscape}). As a consequence, the probability that the system will visit the saddle region is low. These considerations bring us to the notion of entropy, which measures the number of microrealizations that exist for a given macrostate, and is also commonly interpreted as a measure of disorder. As entropy is normally defined for a stable equilibrium state, we define the change in configurational entropy $\Delta S$ with respect to stable fluctuations only \cite{loxley},
\begin{equation}\label{eq:config_entropy}
e^{\Delta S / k_B} \equiv \sqrt{\frac{\beta}{2\pi}} \sqrt{\frac{\prod_i \lambda_i^A}{\prod_j' \lambda_j^S}},
\end{equation}
where $\prod'$ is defined for positive curvatures and an additional $\sqrt{\frac{\beta}{2\pi}}$ factor with $\beta = (k_BT)^{-1}$ is needed to keep the dimension consistent. To clarify our nomenclature, what we previously refer to as the  energy barrier $\Delta E$ is the internal energy barrier, and the total activation energy corresponds to the change in  Helmholtz free energy \cite{loxley}: $\Delta F = \Delta E - T \Delta S$, in which the entropic barrier is given by $- T \Delta S$. It follows that Eq. (\ref{eq:omega0_nozero}) can be expressed as
\begin{equation}\label{eq:new_omega0}
\Omega_0 =   \sqrt{\frac{2\pi}{\beta}} |\lambda_1^S|^{-1/2} e^{\Delta S / k_B},
\end{equation}
in which $\lambda_1^S$ is the negative curvature at $S$. The factor $\Omega_0$ is thus a measure of the number of available configurations and gives the entropic contribution to the prefactor. 
As plotted on Fig. \ref{fig:deltaS}, we find $\Delta S = S_S - S_A < 0 $: the configurational entropy of the metastable state is higher than that of the saddle for all mechanisms considered here. This result implies that the number of micro-realisations of the metastable skyrmion state is higher than that of the transition state. A potential source of stability of individual skyrmions might therefore lie in lowered attempt frequencies due to entropic narrowing in the saddle point region - that is, the existence of an entropic barrier - rather than in topological protection (low internal energy barriers \cite{hagemeister2015stability}). This result was observed experimentally in Ref. \onlinecite{wild2017entropy}, in which Wild and co-workers showed that attempt frequencies in skyrmion lattices are strongly reduced by entropic effects. 
The smaller reduction in entropy between $A$ and $S$ in the boundary annihilation can be explained by the fact that the transition state is a full skyrmion, which remains somewhat similar to the metastable skyrmion state. From Eq. (\ref{eq:new_omega0}), $\Omega_0(i)$ plotted in Fig. \ref{fig:curva_all}c in semilogarithmic scale behaves as $\Delta S(i)$. Therefore, the entropic barrier is primarily associated with the internal modes of a skyrmion.



\paragraph*{}The total rate of annihilation of an individual skyrmion is obtained by the sum of the rates due to collapse and escape through a boundary, $\Gamma_{\text{tot}} =  \Gamma_{\text{col}} + \Gamma_{\text{esc}}$ and, in the absence of defects, remains dominated by $\Gamma_{\text{esc}}$. The escape through a curved boundary is found to be less likely than through a flat boundary, due to both internal energy barrier and entropic barrier being increased. We also stress that the skyrmions in this work are only a few nanometers in radius and stabilized at zero-field. For different stabilization processes involving an external field and lower perpendicular anisotropy, escape processes may be found more favorable also in terms of the internal energy barrier \cite{uzdin2017effect}. Interestingly, the collapse in the presence of another skyrmion exhibits the same internal energy barrier as in the case of a single skyrmion, but a higher entropic barrier, which leads to a lower attempt frequency (see Table \ref{tab:transition_rates}). 
Finally, the presence of a non-magnetic defect significantly affects all terms of the transition rate. Firstly, it lowers the internal energy barrier by almost one order of magnitude, rendering it practically flat (Fig. \ref{fig:Etot_def}). It also decreases the entropic contribution to the prefactor by two orders of magnitude, that is, the entropic barrier is lowered (see also Fig. \ref{fig:deltaS}). Thirdly, the rate of growth of the unstable mode at the saddle point is decreased by one order of magnitude. This effect could, in theory, be stabilizing, but we find it is negligible against the significantly lower value of the total activation energy. It can also be noted that the internal energy barrier for the nucleation is lowered as well. This is consistent with experimental observations that skyrmions tend to nucleate and annihilate near defects \cite{hanneken2016pinning}.



\section{Conclusion and discussion}
\paragraph*{}In the present work, we applied Langer's theory to the computation of annihilation rates of metastable magnetic skyrmions with respect to collapse and escape processes. By changing the underlying symmetry of the DMI, we were able to check that the present results hold for not only Néel skyrmions, but also Bloch skyrmions, and antiskyrmions. We identified the thermally significant modes as the skyrmion's internal modes, while the other modes pertain to collective fluctuations that can be interpreted as spin-wave excitations, and contribute weakly. Additionally, the eigenmodes of saddle configurations exhibit broken symmetries of the metastable modes.

\paragraph*{}Our calculations show that the most probable path to annihilation for a small skyrmion stabilized at zero-field is the collapse on a defect. The presence of a defect seems to significantly lower not only the internal energy barrier, but also the entropic barrier. When no defect is present in the skyrmion's vicinity, escape through a boundary is favoured against collapse, even though it paradoxally involves the highest internal energy barrier and the lowest characteristic growth rate of an instability at the transition state. Therefore, the main source of stability of individual skyrmions in the present system is not found in particularly high internal energy barriers, nor in a slow dynamics at the transition state. Instead, it comes from a narrow saddle region in configuration space, which makes the transition state less likely to be visited under the effect of thermal fluctuations. This result can also be formulated in terms of configurational entropy, which we defined with respect to stable fluctuations: the configurational entropy of the metastable skyrmion state is higher than that of the transition state. This is a case of entropic narrowing in the saddle point region, which leads to lowered attempt frequencies and enhanced stability. This narrowing effect is primarily associated with the skyrmion's internal modes, and is more pronounced for collapse mechanisms. 
As a consequence, we found that due of a lower entropic barrier, and despite a higher internal energy barrier, the escape through a boundary possesses an attempt frequency a thousand times higher than that of collapse.
A curved boundary makes the escape less favorable. Finally, the collapse in the presence of another skyrmion was found to exhibit the same internal energy barrier as the single skyrmion, but a slightly higher entropic barrier. 
The above conclusions highlight the importance of entropic contributions and the necessity to compute a complete activation rate, since the stability of skyrmions cannot be properly assessed solely from estimating internal energy barriers.

\paragraph*{}We used a simple Heisenberg-type model limited to first-neighbor exchange interactions and no dipole-dipole coupling, but we believe it nevertheless captures the essential physics behind skyrmion annihilations. In systems were dipole-dipole interactions were found to play an important role in the skyrmions' stability, it was also demonstrated that an effective anisotropy is enough to reproduce similar energy barriers \cite{lobanov2016mechanism}. Since both activation barriers and attempt frequencies were reported to exhibit a high dependency on external magnetic fields \cite{wild2017entropy}, we can expect that the relative importance of entropic effects is also highly affected by the overall choice of parameters. Contrary to Ref. \onlinecite{bessarab2018annihilation}, in which the translational modes were Goldstone modes, yielding temperature- and size-dependence of the prefactor, here the skyrmions are small compared to the lattice parameter and therefore do not decouple from the lattice, and so the translational invariance is broken by the lattice. Physically, it seems reasonable that Goldstone modes would not contribute in real systems with damping and all the imperfections and interactions responsible for dissipation. 
One common result we share with Ref. \onlinecite{bessarab2018annihilation} seems to be that at low (zero) field, escape is favoured against collapse.


\paragraph*{}Upon identifying the unstable mode at the saddle point, one could imagine suppressing it by strong microwave radiation and thus enhancing stability.
 On the other hand, exciting the internal modes of a skyrmion may bring it over to the saddle point and initiate the collapse. This sort of procedure was previously demonstrated in the case of the melting of a skyrmion lattice by exciting collective modes via an appplied microwave magnetic field \cite{mochizuki2012spin}.

\begin{acknowledgments}
This work was partially supported by the Horizon 2020 Framework
Programme of the European Commission, under Grant agreement
No. 665095 (MAGicSky). Additional support was received from CD-laboratory AMSEN (financed by the Austrian Federal Ministry of Economy, Family and Youth, the National Foundation for Research, Technology and Development), the FWF – SFB project F4112-N13. R. L. S. acknowledges the support of the Natural Sciences and Engineering Research Council of Canada (NSERC). Cette recherche a été financée par le Conseil de recherches en sciences naturelles et en génie du Canada
(CRSNG). We thank Pavel Bessarab for useful remarks concerning the implementation of the GNEB scheme.
\end{acknowledgments}



\appendix


\section{Hessian computation in spherical coordinates}\label{app:hessian}
\paragraph*{}Our system is an assembly of $N$ unit spins $\mathbf{m}=(\hat{m}_1,\dots\hat{m}_N)$ on a lattice. The change in magnitude of the moments is generally much faster than the change in orientation and we can assume that their amplitudes remain constant. The system can thus be described in terms of orientations of the moments alone and the energy surface reduces to a $2N$-dimensional landscape. The total energy may be written in terms of spherical coordinates on the unit sphere $E(\boldsymbol{\theta},\boldsymbol{\phi}) $ where $\boldsymbol{\theta} = (\theta_1 \dots \theta_N)$ is the polar angle with the cartesian Z axis and $\boldsymbol{\phi} = (\phi_1 \dots \phi_N)$ is the corresponding azimuth in the $XY$ plane. In principle, it is necessary to define canonically conjugate variables ($\mathbf{p},\mathbf{q}$) \cite{langer,coffey,seuss} such that 
\begin{eqnarray}\label{eq:conjuguate_var}
\mathbf{p} & = & \cos \boldsymbol{\theta} \nonumber \\  
\mathbf{q} & = & \boldsymbol{\phi}.
\end{eqnarray} 
According to Langer's initial definitions, the total energy is function of $N$ coordinates and $N$ canonically conjuguate momenta \cite{langer}. However, even if the energy is only function of the coordinates, the equipartition theorem holds \cite{duffthesis}. The use of variables such as the ones defined in Eq. (\ref{eq:conjuguate_var}) ensures the Jacobian $J_i= \det \dfrac{\partial(m_{ix},m_{iy},m_{iz})}{\partial(p_i,q_i,1)} = \diff p_i \diff q_i$ is not a function of $(\theta_i,\phi_i)$ \cite{seuss}. In this work, we use spherical coordinates for the computation of the Hessian matrix. This requires corrections in the Hessian to take into account the spherical Jacobian, which we give in what follows. We define the spherical Hessian as:

\begin{equation}\label{eq:sph_hess1}
H = 
\begin{pmatrix}
H_{\theta \theta} & H_{\theta \phi} \\
H_{\phi \theta} & H_{\phi \phi} \\
\end{pmatrix},
\end{equation}
in which
\begin{equation}\label{eq:sph_hess2}
\begin{split}
& H_{\theta_i \theta_j } = \frac{\partial^2 E}{\partial \theta_i \partial \theta_j},
\\
&H_{\theta_i \phi_j } = \frac{1}{\sin \theta_j}\frac{\partial^2 E}{\partial \theta_i \partial \phi_j}, \\
& H_{\phi_i \theta_j } = \frac{1}{\sin \theta_i}\frac{\partial^2 E}{\partial \phi_i \partial \theta_j},\\
&H_{\phi_i \phi_j } = \frac{1}{\sin \theta_i \sin \theta_j}\frac{\partial^2 E}{\partial \phi_i \partial \phi_j}. 
\end{split}
\end{equation}
Even though the total Hessian remains symmetric, it is necessary to remain cautious with the introduction of a DMI contribution as it makes the $H_{\theta \phi}$ submatrix non symmetric. Therefore in general, $H_{\phi \theta} = H_{\theta \phi} ^T \neq
H_{\theta \phi}$, contrary to the way it was treated in \onlinecite{coffey}.


\section{Derivation of the transition matrix of LLG}\label{app:LLG}
The dynamical prefactor takes into account the dynamics of the system at the saddle point and is derived from the set deterministic Landau-Lifshitz-Gilbert (LLG) equations associated with each spin $\hat{m}_i, i =1..N$\cite{coffey},
\begin{equation}\label{eq:LLG}
\frac{\diff \hat{m}_i}{\diff t} = - \Big[ g' \hat{m}_i \times \frac{\partial E}{\partial \hat{m}_i } + h' \big( \hat{m}_i \times \frac{\partial E}{\partial \hat{m}_i } \big) \times \hat{m}_i \Big],
\end{equation}
where
\begin{equation}
g' = \frac{\gamma}{(1+\alpha^2)M_s}
\end{equation} 
corresponds to the gyromagnetic ratio $\gamma$ modified by a dimensionless damping factor $\alpha = \eta \gamma M_s$, in which  $M_s$ is the saturation magnetization and $\eta$ is a damping parameter characterizing the coupling to the heat bath, and 
\begin{equation}
h' = \alpha g'.
\end{equation}
It follows that the first term on the RHS of Eq. (\ref{eq:LLG}) is the Larmor equation describing the precession of the magnetization vector $\hat{m}_i$, and the second term is an alignment term whose effect is measured by $h'$. Re-formulating Eq.  (\ref{eq:LLG}) within the local spherical basis $(\hat{e}_r,\hat{e}_{\theta},\hat{e}_{\phi})$ yields the following set of differential equations:
\begin{eqnarray}\label{eq:sph_LLG}
\dot{\theta}_i & = & \frac{g'}{\sin \theta_i} \frac{\partial E}{\partial \phi_i} - h' \frac{\partial E}{\partial \theta_i}, \nonumber  \\
\dot{\phi}_i & = & \frac{-g'}{\sin \theta_i} \frac{\partial E}{\partial \theta_i} - \frac{h'}{\sin^2 \theta_i} \frac{\partial E}{\partial \phi_i}.
\end{eqnarray}

The next step consists in approximating the energy close to the saddle point $(\tilde{\boldsymbol{\theta}},\tilde{\boldsymbol{\phi}}) =  (\tilde{\theta_1}, \dots \tilde{\theta_N}, \tilde{\phi_1}, \dots \tilde{\phi_N})$ as a Taylor series truncated to the second order term (Eq. (\ref{eq:taylorseries})), followed by a derivation of the obtained expression with respect to $(\theta_i,\phi_i)$. Finally, setting $\boldsymbol{\Theta} = \boldsymbol{\theta} - \tilde{\boldsymbol{\theta}} $ and $\boldsymbol{\Phi} = \boldsymbol{\phi} - \tilde{\boldsymbol{\phi}} $, (\ref{eq:sph_LLG}) reduces to the following system of equations linearized about the saddle point, which we write in matrix form,
\begin{equation}
\begin{pmatrix}
\dot{\boldsymbol{\Theta}} \\
\\
\dot{\boldsymbol{\Phi}} \\
\end{pmatrix}
=
\begin{pmatrix}
\mathcal{T}_{\theta \theta} & \mathcal{T}_{\theta \phi} \\
\mathcal{T}_{\phi \theta} & \mathcal{T}_{\phi \phi} \\
\end{pmatrix}
\begin{pmatrix}
\boldsymbol{\Theta} \\
\\
\boldsymbol{\Phi} \\
\end{pmatrix},
\end{equation}
where
\begin{equation}
\begin{split}
&\mathcal{T}_{\theta_i \theta_j} = g' H_{\phi_i\theta_j}^S -h'H_{\theta_i\theta_j}^S, \\
&\mathcal{T}_{\theta_i \phi_j} =  \sin \tilde{\theta}_j \big(g' H_{\phi_i\phi_j}^S - h'H_{\theta_i\phi_j}^S \big), \\
&\mathcal{T}_{\phi_i \theta_j} = -\frac{1}{\sin \tilde{\theta}_i} \big( g' H_{\theta_i\theta_j}^S + h' H_{\phi_i\theta_j}^S  \big), \\
&\mathcal{T}_{\phi_i \phi_j} = -\frac{\sin \tilde{\theta}_j}{\sin \tilde{\theta}_i} \big(g' H_{\theta_i\phi_j}^S + h' H_{\phi_i\phi_j}^S  \big),
\end{split}
\end{equation}
define the submatrices of the transition matrix and $H_{\theta \theta}^S$, $H_{\phi \phi} ^S$, $H_{\theta \phi} ^S$ ,$H_{ \phi \theta}^S$  are the submatrices in the spherical Hessian defined in Eq. (\ref{eq:sph_hess2}) and evaluated at $S$.


\section{Atomistic simulations} \label{app:simulations}
\paragraph*{}We simulate a strictly two-dimensional surface reprensenting a thin magnetic layer.  The total simulated domain contains $50 \times 50$ spins or $80 \times 80$ spins in the configuration involving two skyrmions, and we keep open (non-periodic) boundary conditions. Néel-type skyrmions are stabilized by interfacial DMI for which the Dzyaloshinskii vector is  defined as $\vec{D_ {ij}} = D \hat{r}_{ij} \times \hat{e}_z$, where $\hat{r}_{ij} $ is the in-plane direction between sites $i$ and $j$ \cite{thiaville2012dynamics}. Additionally, Bloch-type skyrmions are stabilized by bulk-type DMI with $\vec{D_ {ij}} = D \hat{r}_{ij}$ \cite{butenko}, while antiskyrmions are favoured by modifying the interfacial DMI such that  $\vec{D_ {ij}} = - D \hat{r}_{ij} \times \hat{e}_z$  when $\hat{r}_{ij}=\hat{e}_y$ \cite{ritzmann2018trochoidal}. A non-magnetic defect is simulated by setting the anisotropy to zero at a given site as well as all exchange interactions with neighboring sites. Similarly, a disk-shaped sample is obtained by converting outer spins into non-magnetic sites.  We use an isotropic exchange constant of $J_{\text{ex}} = 1.6\times 10^{-20}$ J ($\sim 100$ meV) with lattice constant $a=1$ nm and saturation magnetization $M_s = 1.1a^3 \times 10^6$ A.m$^{2}$  \cite{thiaville2012dynamics}. The chosen values of the parameters allow for the stabilization of small, individual, metastable skyrmions at zero-field with a radius of approximately ten lattice sites \cite{heo2016switching}, and are the following: $D/J_{\text{ex}} = 0.36$, $K/J_{\text{ex}} = 0.4$. The damping term in the LLG Eq. (\ref{eq:LLG}) is set to $\alpha = 0.5$, which corresponds to commonly found values for ultrathin magnetic films with DMI \cite{thiaville2012dynamics}, while pertaining to the IHD regime. The gyromagnetic ratio is that of the free electron, $\gamma = 1.76 \times 10^{11}$ rad.s$^{-1}$.T$^{-1}$. The  CI-GNEB scheme is used on $Q=10$ or 15 images of the system for single-skyrmion mechanisms, and $Q=6$ images for the two-skyrmion mechanism. 
Our implementation of the Hessian was analytical to minimize numerical noise as much as possible. In case one or several spins lie at the pole of the sphere at either $A$ or $S$, the whole system is rotated in order to avoid the singularity of the spherical coordinate system. The computation of the complete activation rate was tested against the analytical formula derived in Ref. \onlinecite{schratzberger2010validation} Eq. (24) of the attempt frequency of magnetization reversal for a single macro-spin in a perpendicular field. We simulated a 0.6 $\times$ 0.6 nm$^2$ sample and reproduced with a good agreement Fig. 1a of Ref. \onlinecite{seuss}.

\subsection*{Influence of the lattice size.} As was discussed in Ref. \onlinecite{seuss}, the accuracy of $\Omega_0$ can decrease significantly due to potential numerical errors being multiplied. In order to check our results, we computed the activation rates on a lattice of various sizes. The results are gathered in Table \ref{tab:lattice_sizes}. For processes involving a single skyrmion on 30 $\times$ 30 simulated sites, the skyrmion is constrained by the boundary and less stable. However, at 50 $\times$  50 sites and above, we observe very little variations in the different terms of the prefactor, which increases confidence in the present results. In the case of the two-skyrmion process, the rates loses its lattice-size dependency for $70 \times 70$ simulated sites and above. 

\begin{table}
\centering
\caption{Terms of the rate prefactor calculated for different number of sites $N = N_x \times N_y$ (here $N_x = N_y$).\label{tab:lattice_sizes}}
\subcaption{Collapse.}
\begin{ruledtabular}
\begin{tabular}{ccccc}
$N$ & $\Delta E$ ($J_{\text{ex}}$) & $\Omega_0  (\times 10^{-5})$ &  $\lambda_+$ (GHz) & $ \Gamma_0$ (MHz)\\
\hline
 $30\times30$ & 2.76 & 3.79 & 1198.73 & 7.24 \\
 $50\times50$ & 2.83 & 3.51& 1200.47 & 6.70 \\
 $70\times70$ & 2.83 & 3.49 & 1200.48 & 6.67 \\
\end{tabular}
\end{ruledtabular}
\bigskip
\subcaption{Escape (flat boundary).}
\begin{ruledtabular}
\begin{tabular}{ccccc}
$N$ & $\Delta E$ ($J_{\text{ex}}$) & $\Omega_0  (\times 10^{-2})$ &  $\lambda_+$ (GHz) & $ \Gamma_0$ (GHz)\\
\hline
$30\times30$ & 3.23 & 1.20 & 521.6 & 3.80 \\
$50\times50$ & 3.28 & 1.24 & 522.94 & 4.14 \\
$70\times70$ & 3.28 & 1.24 & 522.94 & 4.13  \\
\end{tabular}
\end{ruledtabular}
\bigskip
\subcaption{2 sk. collapse.}
\begin{ruledtabular}
\begin{tabular}{ccccc}
$N$ & $\Delta E$ ($J_{\text{ex}}$) & $\Omega_0  (\times 10^{-5})$ &  $\lambda_+$ (GHz) & $ \Gamma_0$ (MHz)\\
\hline
$50\times50$ & 2.82 & 1.86 &  1198.11 & 3.56 \\
$60\times60$ & 2.82 & 2.49 & 1200.22 &  4.76 \\
$70\times70$ &  2.82 &  2.33 & 1200.23 & 4.44\\
$80\times80$ & 2.82 & 2.32 & 1200.23 & 4.43 \\
\end{tabular}
\end{ruledtabular}
\end{table}

\newpage
\bibliography{/home/louise/Dropbox/writing/latex/skyrmionbib}

\end{document}